\documentclass[11pt]{article}
\usepackage[multiple]{footmisc}

\usepackage{macros}
\usepackage{threeparttable}
\usepackage{pdflscape}
\usepackage{afterpage}

\usepackage{clipboard}
\newclipboard{draft}

\newcommand{\obs}{\text{obs}}
\newgeometry{margin=1in}
\title{JUE Insight: The (Non-)Effect of \\ Opportunity Zones on Housing
Prices\thanks{Chen:
\href{mailto:jchen@hbs.edu}{ \texttt{jchen@hbs.edu}}; Glaeser:
\href{mailto:eglaeser@harvard.edu}{ \texttt{eglaeser@harvard.edu}}; Wessel:
\href{mailto:DWessel@brookings.edu}{ \texttt{dwessel@brookings.edu}}. 
We thank the editor and two anonymous referees for helpful comments that improved the
paper substantially.
We thank
participants of AEA 2021 session ``Do OZs Create Opportunities''
and the Brookings Institution conference on OZs, as well as Isaiah
Andrews, Fernando Ferreira, Amy Finkelstein, Amit Khandelwal, Mike Luca,
Ashesh Rambachan,
Jonathan Roth, Suproteem Sarkar, Liyang Sun, Susan Wachter, and Christopher
Walker for helpful discussions. We especially thank Yanchen Jiang for superb
research assistance. All errors are our own. All replication files are available
at \url{https://github.com/jiafengkevinchen/OZ-CGW}.}}
\author{Jiafeng Chen \\ Harvard University \and
Edward Glaeser \\ Harvard University \and David Wessel \\ Brookings
Institution}

\begin{document}

\begin{titlepage}
    \maketitle
    \begin{abstract}
    Will the Opportunity Zones (OZ) program, America’s largest new place-based policy in
    decades, generate neighborhood change? We compare single-family housing price growth
    in OZs with price growth in areas that were eligible but not included in the
    program. We also compare OZs to their nearest geographic neighbors. \Copy{abs}{Our most credible estimates
        rule out price impacts greater than 0.5 percentage points with 95\% confidence},
    suggesting that, so far, home buyers don't believe that this subsidy will generate
    major neighborhood change. OZ status reduces prices in areas with little employment,
    perhaps because buyers think that subsidizing new investment will increase housing
    supply. Mixed evidence suggests that OZs may have increased  residential permitting.
    \end{abstract}
\end{titlepage}

\section{Introduction}
\label{sec:intro}

Opportunity Zones (OZs), established by the Tax Cuts and Jobs Act of 2017
(TCJA), arguably represent the U.S. government's largest place-based policy
innovation since Empowerment Zones were introduced in 1993.    When capital
gains are invested in OZs, taxes on the original gains are deferred and can be
reduced; taxes on future gains from qualifying investments in OZs are largely
eliminated.   Will this significant tax-based subsidy lead to neighborhood
improvements?

\Copy{citehouseshock}{We test this hypothesis by examining whether areas
that are designated as OZs
subsequently experience an increase in residential real estate prices. If OZ designation
is seen as a harbinger of future investment, then new buyers should presumably
anticipate future neighborhood improvements and be willing to pay more for homes.  If
OZs are ineffective, or act primarily to generate more residential supply, then OZs will
have little impact on price. Indeed, there is some empirical evidence that
local
housing prices respond to changes in neighborhood-level housing supply; see
\cite{asquith2019supply,pennington2021does,li2019new}}.  It is too early to test the
more interesting and important question of whether OZs impact people as well as place,
and whether they positively impact the lives of neighborhood residents.\footnote{%
\cite{busso2008local} provide a thorough analysis of the economic impact of
national employment zones, which appear to have meaningfully impacts both
housing prices and employment.    \cite{neumark2010enterprise} find fewer
positive effects of state level Enterprise Zones. } Moreover, we only have price
data available for 2018--2020, and the OZ tracts were designated by governors in
the first few months of 2018 and officially posted in July.%

A non-peer reviewed study of housing prices, done by Zillow \citep{zillow}, found a
positive effect of OZs on prices when comparing OZ areas with areas that did not receive
zone status, but there are two reasons to be cautious about this work.  First, the
pre-2017 trends in prices between their treatment and control samples do not appear to
be parallel. They provide no tested pre-trends. Second, \cite{zillow} results also do
not control for changes in the quality of houses sold, since they include all arms'
length sales prices. Relatedly, another non-academic study by \cite{attom} observes that
OZs found median home price increases, also using proprietary data. However, the study
does not compare OZs to similar places without OZ designation, and thus its results
cannot be interpreted causally. More recently, the \cite{cea} replicated our analysis
from the working paper version of this paper \citep{NBERw26587}.

We use the FHFA repeat sales-indices for
single-family homes at the tract and ZIP code level to measure price
changes \citep{bogin2019local}.\footnote{Our data do not include multi-family
housing or commercial
real estate. See \cite{sage2019opportunity} on possible impact of OZs on prices of commercial real estate and land.} We perform three different
empirical exercises. First, we follow  \cite{zillow} and
compare OZ areas with areas that were initially eligible for
OZ status, but then not included as OZs, assuming
and testing for parallel trends. Second, we use propensity-score weighting
methods to include observed characteristics nonparametrically in the
difference-in-difference design, making and testing a conditional parallel
trends assumption. Third, we compare OZ areas with bordering
areas. As many tracts have missing data, we perform the exercises at the ZIP
code level in addition to at the tract level as well.

All exercises yield a similar result:  OZs appear to have a
negligible price impact that is statistically indistinct from zero.   Our
results are sufficiently precise that we can generally rule out a
tract-level price impact
of $1.5$ percentage points or more annually. Our point estimates are typically between $0$ and $0.5$ percentage
points. This finding suggests that buyers do not believe that OZ
status will generate a significant change in the economic fortunes of the
neighborhood. Alternatively, buyers could be  myopic, but that seems unlikely if
the zone status attracts professional investors. One possible explanation for
our null result is that the overall treatment effect combines a positive shock
to housing demand and a positive shock to housing supply, since OZ
status provides financial incentives that could encourage more residential
construction. To test this possibility, we compare ZIP codes in predominantly
residential areas with ZIP codes that are predominantly commercial areas. We
hypothesized that the positive impact of OZs on housing supply
should be stronger in already residential areas, and so the impact of
OZ status on housing prices should be lower in those areas.

With an interaction specification, we do find that the treatment effect of zone
status is weakly positive in non-residential areas, and the interaction between
OZ status and being predominantly residential is negative.   Our
point estimates do indeed suggest that OZs appear to have
increased prices in less residential areas and reduced prices in more
residential areas.  Yet the estimated effects are small and our standard
errors are too large to rule out a zero impact in either type of area.

We supplement our primary work on prices by examining the impact of OZs on building permits, which is our best measure of new housing supply.
Since places are a much coarser geography, we present results using a simple
difference-in-difference methodology.  We do find a small positive impact of
OZ on the number of permits, but not on the number of permits
relative to the stock of housing in 2010. Our results are compatible with the
view that OZ status encouraged some construction, especially in
smaller places.

\Copy{othereffects}{This work does not imply that OZs were a mistake or
that there are no benefits from
these zones. These tax subsidies may generate neighborhood change in the future that
buyers do not anticipate today.\footnote{There is also mixed evidence on potential
employment benefits; see 
\cite{arefeva2021job,atkins2020impact,freedman2021jue}.} The
costs of these subsidies may end up being so small that they are offset by even tiny
price gains. Nonetheless, the absence of a visible price effect does suggest the limits
of place-based policies, especially those that focus on investment in physical rather
than human capital.}

This paper proceeds as follows. \Cref{sec:inst} discusses the institutional context,
policy details, and selection procedures for OZs. \Cref{sec:data} discusses the various
data sources used in our study. \Cref{sec:strat} introduces our three main empirical
strategies. \Cref {sec:res} discusses the empirical results. \Cref{sec:conc} concludes.
We relegate a description of various data sources used in the study to \cref{sec:data}.

\section{Institutional Context}
\label{sec:inst}
OZs were created in December 22, 2017, when President Trump signed
the TCJA into law. They
are intended to spur economic development in distressed communities. The law
provides three benefits for investing capital gains in one of 8,762 Census
tracts\footnote{Two additional tracts in Puerto Rico were added by subsequent
legislation, bringing the total to 8,764. We do not have housing data for Puerto
Rico.} (12\% of all tracts) across the country through intermediaries
called Opportunity Funds: Tax on the initial capital gain is deferred until 2026
or when the asset is sold. For capital gains placed in an Opportunity Fund for
at least five years, investors’ basis on the original investment increases by
10\%; if invested for seven years, by 15\%.\footnote{The 10\% step-up in basis
is only for OZ investments made by Dec. 31, 2021. The additional
5\% step-up in basis is for investments by Dec. 31, 2019.} For investments in
Opportunity Funds held for at least 10 years, the gains on the investments in
the zones are not subject to capital gains tax. Funds can be invested in
commercial, residential and industrial real estate; infrastructure, and
businesses. For real estate projects to qualify, the investment must result in
the property being ``substantially improved.''

    The outlines of a proposal to create OZs was published by the
 Economic Innovation Group in April
 2015 \citep{eig}.
 Bills to create them were first introduced in Congress in April 2016 and
 re-introduced in February 2017, but got little attention initially. As in the
 bill that eventually became law, the proposals authorized governors to nominate
 as OZs 25\% of the ``low income communities'' in their states; in
 states with fewer than 100 low income communities, the governor could choose
 25. The definition of low income communities was borrowed from a 2000 law that
 created the New Markets Tax Credit: tracts were designated as low income
 if the poverty rate is at least 20\%, or  the median family income doesn’t
 exceed 80\% of the statewide median for a tract outside the metropolitan area,
 or the median income doesn’t exceed 80\% of the statewide median or the metro
 area statewide median for a tract inside a metropolitan area. Tracts contiguous
 with low income communities also are eligible, provided their median family
 income doesn’t exceed 125\% of the contiguous low income community.
 
    The OZ provision was not included in the House version of the
    TCJA, which was introduced on Nov. 2, 2017.  With very little public
    attention, Sen. Tim Scott (R, S.C.), a member of the Senate Finance
    Committee, successfully pushed to include the OZ provision in
    the TCJA, which was introduced on Nov. 28, 2017.  The first reference in the
    press to the OZ provision in the TCJA came on November 28,
    2017, in South Carolina’s Post and Courier, according to the Factiva
    database \citep{lovegrove}.
      
    Governors had 90 days after the passage of the law---until March 21, 2018
    unless they sought a 30-day extension---to nominate zones from a list of
    31,866 eligible tracts prepared by the Treasury based on 2010-2015
    American Community Survey data.  The Treasury posted a list of all qualified
    OZs on July 9, 2018.   Of the 8,764 OZs, 8,534
    are low income communities and 230 are contiguous tracts. A map
    of the zones in the U.S. mainland is shown in \cref{fig:map} in the
    appendix.
    
\section{Data}
    \label{sec:data}
    Our measure of housing price growth is the annual change in the housing
    price index computed by the Federal Housing Finance Agency (FHFA). The
    housing price index is a weighted, repeat-sales price index of the movement
    of single-family house prices. Like all repeated-sales
    indices,
    it
    attempts to correct for quality changes. We use the annual house
    price indices by tract and by five-digit ZIP 
    codes,\footnote{Links for data sources are provided in Appendix \cref{tab:data_sources}, accessed as of March 18, 2022.} and treat 2018 and 2019
    as the treated years.

Information about the OZs is provided by the Urban
Institute.
The data includes whether a tract belongs to the 31,866 eligible tracts and to the
selected 8,762 tracts and whether a tract is eligible for selection as a low income
community or as contiguous tracts, which we use in our first and second empirical
designs in \cref{sec:strat}. Characteristics of the Census tracts are from the American
Community Survey (ACS) 2013--2017 5-year estimates. \Copy{sampleinfo}{Among the 8,534
low-income selected tracts, 3,806 have price information in the sample period
(2014--2020), of which only 2,917 have price information for the entire sample period.
We discuss the representativeness of the various subsamples in
\cref{asub:unbalanced,asub:represent}, and our results are robust to
using balanced versus unbalanced panels.} Employment data at the tract
level are available in the Longitudinal Employer-Household Dynamics
data.
    
Geographical comparison between tracts and their non-selected geographical
neighbors uses the TIGER 2018 geographic shapefiles provided by the
Census.
Aggregating tract-level data to ZIP-level data, implemented in 
\cref{sub:zip}, uses the
geographical crosswalks between 5-digit USPS ZIP codes and Census
tracts are provided by the Office of Policy Development and Research at
the Department of Housing and Urban Development, where we use the data
for the first quarter of 2017.
Lastly, splitting ZIP codes by employment population in 
\cref{tab:hetero_effects_} uses
population data at the ZCTA level in the 2012--2016
ACS 5-year estimates and employment data from the ZIP code level County
Business Patterns in 2016.

We use the Building Permits Survey (BPS) for data on
residential
permitting. The finest geographical level that the BPS reports is the
Census place, which is non-overlapping with tracts. As a result, we
aggregate tracts up to places via crosswalk provided by the Missouri Census
Data Center, which
also provides estimates for housing units in 2010. Since New York City's
boroughs are different Census places but have the same FIPS identifier, we
drop New York City from our analysis.

\section{Empirical Strategy}
\label{sec:strat}

We use three empirical strategies for the tract-level analyses. First,
following \cite{zillow}, we compare OZs to tracts that are eligible for
OZs but are not selected.\footnote{In the results presented in the main text, we
compare selected tracts to tracts that are eligible but not selected
\emph{conditional on} the eligibility criterion being \emph{low-income
community}, as tracts that are eligible for contiguity reasons are
overrepresented in the non-selected group. Qualitative conclusions do not change
if we remove this condition (\Cref{asec:all_oz}).} In this case, we use a
difference-in-differences design that optionally incorporates observable
tract-level covariates interacted with year fixed effects. We supplement the
analysis in \cite{zillow} with formal tests of pre-treatment trends. Second, we
refine our analysis in the first design with semiparametric propensity score
weighting methods in a difference-in-differences setting
\citep{abadie2005semiparametric,callaway2018difference,sant2018doubly}. Third,
we compare OZs with their geographical neighbors that are not selected.

The FHFA tract-level data covers only about half of the selected tracts. Concerned with
the data attrition, we also aggregate tract-level data to the ZIP code level, and uses
FHFA data at the ZIP code level, which has better
coverage.\footnote{\Copy{footzip}{        Although only $17,161$ of
        the total $39,300$ ZIP codes with crosswalk data
        do not have missing data in 2018,
        these ZIP codes intersect with
        $6,988$ selected Opportunity Zones.%
}} We
split the ZIP-code level data
into quantiles by employment population so as to decompose the potential effects of OZs
on  supply and demand, noting that the positive supply effect has larger impact in
residential areas. Lastly, we perform a similar analysis at the Census place level for
the effect on permitting.

\Copy{notation}{In the following description of methodology, we let $Y_
{it}$ denote
the observed outcome
in geography $i$ and time $t$. We let $D_i$ denote OZ status and $D_{it}$
denote the interaction between OZ status and post-treatment time. We
assume that the data across geography are i.i.d. sampled. As in the standard setting,
the estimate target is the average treatment effect on the treated (ATT),
mean-aggregated over post-treatment periods.}

\subsection{Baseline difference-in-differences \citep{zillow}}
\label{sub:twfe}
In the first design, we compare tracts that are selected as OZs to tracts that
are eligible but not selected. The parallel trends assumption allows us to
identify the treatment effect in a difference-in-differences design. As we see
in \cref{fig:zillow_event_study}(a), parallel trends is a more plausible
assumption when comparing selected tracts  to eligible but not selected tracts
than when comparing selected tracts to all other tracts.

Our first estimation strategy uses a specification where we allow
for tract-specific heterogeneity in levels and overall trends
\citep{angrist2008mostly} via two-way fixed effects (TWFE): $Y_{it}(0) =
\mu_i +
\alpha_t + \epsilon^0_{it}$ and $
Y_{it}(1) = \mu_i + \alpha_t + \tau + \epsilon^1_{it}$
with $\E[\epsilon_{it}^0 \mid D_i, \mu_i] = \E[\epsilon_{it}^1
\mid D_i, \mu_i] = 0$.\footnote{The model assumes constant treatment effects.
When treatment effects are heterogeneous, the TWFE estimator is
consistent for a weighted average of individual treatment effects.} The usual
parallel trends assumption is implied by the outcome model, since $\E[Y_ {it}(0)
- Y_{i,t-1}(0) \mid D_i = d, \mu_i] =
\alpha_t - \alpha_{t-1}$, which does not depend on treatment status $d$.
Thus $\tau$ is consistently estimated by $\hat \tau$ in the panel OLS
regression \begin{equation}
\label{eq:twfe}
    Y_{it}^\obs = \mu_i + \alpha_t + \tau \one(t \ge t_0, D_i=1) +
    \epsilon_{it}
\end{equation}

\subsection{Propensity-score-weighted difference-in-differences}
\label{sub:santanna}

Identification using the selection of OZs from eligible Census
tracts faces the challenge that, selected tracts and tracts that
are eligible but not selected differ in observable characteristics. Unbalanced characteristics suggests
that parallel trends may not hold.\footnote{%
We do find substantial imbalance in covariate values in
\cref{tab:balance_of_selected_opportunity_zones_and_eligible_census_tracts,tab:paired_balance}
in the appendix. Technically, the identification assumption, (conditional)
parallel trends, does not require covariate balance, since trends are only
required to be parallel and may differ in levels. } We tackle this
challenge with a propensity-score weighting approach \citep{abadie2005semiparametric}. Recent work by
\cite{callaway2018difference} extends 
\cite{abadie2005semiparametric} to settings with multiple periods, multiple
treatment groups, and multiple treatment timings.\footnote{In our design, we
only have multiple pre-periods. There is a recent literature on the failure of
the two-way fixed effect estimator \eqref{eq:twfe} in situations with variable
treatment timing and heterogeneous treatment effects, as the estimator becomes a
weighted average of individual treatment effects with non-convex weights in
large samples
\citep{abraham2018estimating,athey2018design,callaway2018difference,goodman2018difference,imai2016should,borusyak2017revisiting,de2017fuzzy}.
The issue is not as pertinent in our setting as we do not have variable
treatment timing.} 
\cite{sant2018doubly} extend
\cite{abadie2005semiparametric}'s two-period, two-group model and introduces a
doubly-robust version of the semiparametric estimator in
\cite{abadie2005semiparametric}. The ATT estimator is consistent if either the
propensity score model or the outcome regression model is correctly specified.

\subsection{Comparison of geographical neighbors}
\label{sub:pairs}

\Copy{whypaired}{Our third empirical strategy uses geographic proximity to construct a
paired sample of tracts. One may be concerned that certain unobserved, possibly
time-varying, confounders drive selection of OZs---endangering causal interpretation of
comparisons between OZs and their eligible-but-not-selected counterparts. If these
unobserved factors affect geographically close regions similarly, then comparing OZs
with their non-OZ geographical neighbors yields a more plausible natural experiment. On
the other hand, geographical comparisons fall prey to potential spillover effects: If OZ
selection has positive spillover effects,\footnote{For instance,
\citet{hanson2013spatially}finds evidence that the Empowerment Zone program has
significant spillover effects. Investigation of spillover effects is beyond the scope of
this paper, but recent methodological work by \citet{butts2021difference} may further
shed light.} then comparisons with geographical neighbors understate treatment effects,
and vice versa.}

For each selected tract $i$, we construct its non-selected {\emph{neighbor}}
$\tilde i$ to be the tract that is (i) not selected as an OZ, (ii) closest to $i$ by
distance between centroids, (iii) in the same state as $i$, and (iv) has no missing
housing price data in 2018.\footnote{The distance between centroids is calculated using
the Haversine formula, assuming the Earth is a sphere with radius 6,371 kilometers.
The average centroid distance between pairs             is $2.662$             ($5.695$) kilometers.} Within each pair $\iota = (i,\tilde i)$, we
specify $Y_{it}(0) = \gamma_\iota +
\alpha_{\iota t} + \epsilon_{\iota t}$, $Y_{\tilde it}(0) = \tilde \gamma_\iota
+ \alpha_{\iota t} + \tilde
\epsilon_
{\iota t}
$ such that $Y_{it}(0) - Y_{\tilde it}(0) = (\gamma_\iota - \tilde
\gamma_\iota) + (\epsilon_{\iota t} - \tilde\epsilon_{\iota t})$, leading
to the estimation procedure \begin{equation}
\label{eq:paired}
    Y_{it}^\obs - Y_{\tilde it}^\obs = \tau \one(t \ge t_0, D_i = 1) +
    \mu_\iota +
\eta_{\iota t},
\end{equation}
which is consistent assuming $\E[\eta_{\iota t} \mid D_i, \mu_\iota] = 0$. The
identification assumption in the third strategy is the pair, $i, \tilde i$ has
the same trend $\alpha_{\iota t}$, which is differenced away as we construct the
estimator. 

\subsection{Heterogeneity analyses}
\label{sub:hetero_res_method}

\Copy{hetero}{We also use employment data from Longitudinal Employer-Household Dynamics
Origin-Destination Employment Statistics (LODES) to classify tracts as either
residential or commercial.\footnote{Precisely speaking, we use the total non-federal
employment population who works at the census tract in 2017, to match our population
data from the 2017 ACS (WAC All Jobs Excluding Federal Jobs).}
We define residential tracts as those with the employment-to-population ratio
(population employed in the tract to population residing in the tract) being below
median.  We then perform the analysis of \cref{sub:twfe}, interacting the design with
the residential indicator, in order to probe the potential heterogeneous effects of OZ
designation. In a similar vein, we also conduct this heterogeneity exercise
by classifying tracts as either having high or low price elasticity of
housing supply, per the estimates of \cite{baum2019microgeography}.
}

\subsection{Aggregating to ZIP codes}
\label{sub:zip}
The FHFA tract-level data only covers less than half of all treated OZs.
Moreover, the sample suffers from further attrition due to panel balance and
missing Census covariates. To address the data availability concern, we include
an alternative design by aggregating tracts to the ZIP code level. Mimicking our
tract-level analysis, we drop ZIP codes that do not intersect with any Census
tracts that are eligible to be selected as OZs.\footnote{Our
empirical results are not sensitive to this choice.} Each ZIP code $z$ is
partitioned into tracts $i \in I_z$, with $\pi_i^z$ proportion of total
addresses within tract $i$. We choose this aggregation method because crosswalks
are readily provided by the Department of Housing and Urban Development and
addresses are the most relevant measure for residential housing prices. For each
variable $V$, we construct $V_z = \sum_{i \in I_z} \pi_i^z V_i.
$
The ZIP-code level treatment exposure $D_z$ now has a continuous distribution on
$[0,1]$. The analysis then proceeds as in the tract case. For weighting based estimator
in \cite{sant2018doubly}, we discretize $D_z$ by taking $\tilde D_z = \one (D_z \ge q)$
where $q$ is chosen such that the sample mean of $\tilde D_z$ equals the proportion of
treated tracts among all tracts.%

\subsection{Effect on residential permitting}
\label{sub:permitting}

\Copy{placedef}{We study the effect of OZs on residential permitting in a similar
manner. Our data for residential permitting, the Building Permits Survey (BPS), reports
at the Census place level; therefore, we aggregate tracts up to Census places in a
similar manner as in \cref{sub:zip}.

We define treatment as the binary variable which is one if a Census place intersects
with a low-income community OZ, and zero otherwise; we restrict the sample to places
that intersect with any eligible tract.\footnote{We have permit data for 7,241 Census
places, where 3,613 intersect with eligible Census tracts.} About 50\% of the sample are
classified as treated, covering 5,162 OZs; the median treated place has 22\% of its
housing units falling into an OZ. In parts of our analysis, we also use a linear
specification\footnote{The specification is entirely analogous to our TWFE analysis of
ZIP-level aggregated data.} with the treatment intensity, defined as the proportion of
housing units in a place falling into a low-income community OZ, as a robustness check.

The BPS reports permitting data in terms of units, buildings, or dollar value, and, for
each measure, the BPS reports by housing type (single family vs. multifamily, etc.). We
use total\footnote{i.e., aggregating over all housing types; in \cref{asub:unittype}, we
decompose outcome variables into unit types.} units and value as our main variables of
interest, and sum over different housing types. }

We then perform a difference-in-difference analysis comparing treated places to
untreated places. The building permits data is at a monthly level, which we aggregate
every three months to smooth out some of the noise. We define December 2017 as the
treatment time.\footnote{\Copy {treatmenttime}{The choice corresponds to the passage of
the TCJA and hence a reasonable estimate of the earliest time point where anticipation
of treatment could occur; doing so also makes the permitting results comparable to the
price results. Moreover, changing the treatment time would only change the baseline in
the event study plot in \cref{fig:quant}.}}
\Copy{cic}{Since there are many
choices of the outcome variable (e.g., levels vs. logs), we supplement our
analysis with a changes-in-changes
\citep{athey2006identification} analysis in \cref{asec:cic} that looks at
the distribution of outcomes.\footnote{Of course, the assumptions required
for differences-in-differences for different outcome variables and for
changes-in-changes are different \citep{roth2020parallel}.}}

\begin{figure}[tb]
\centering

(a) Raw trend of housing prices by treatment status
\vspace{-1em}
\includegraphics[width=\textwidth]{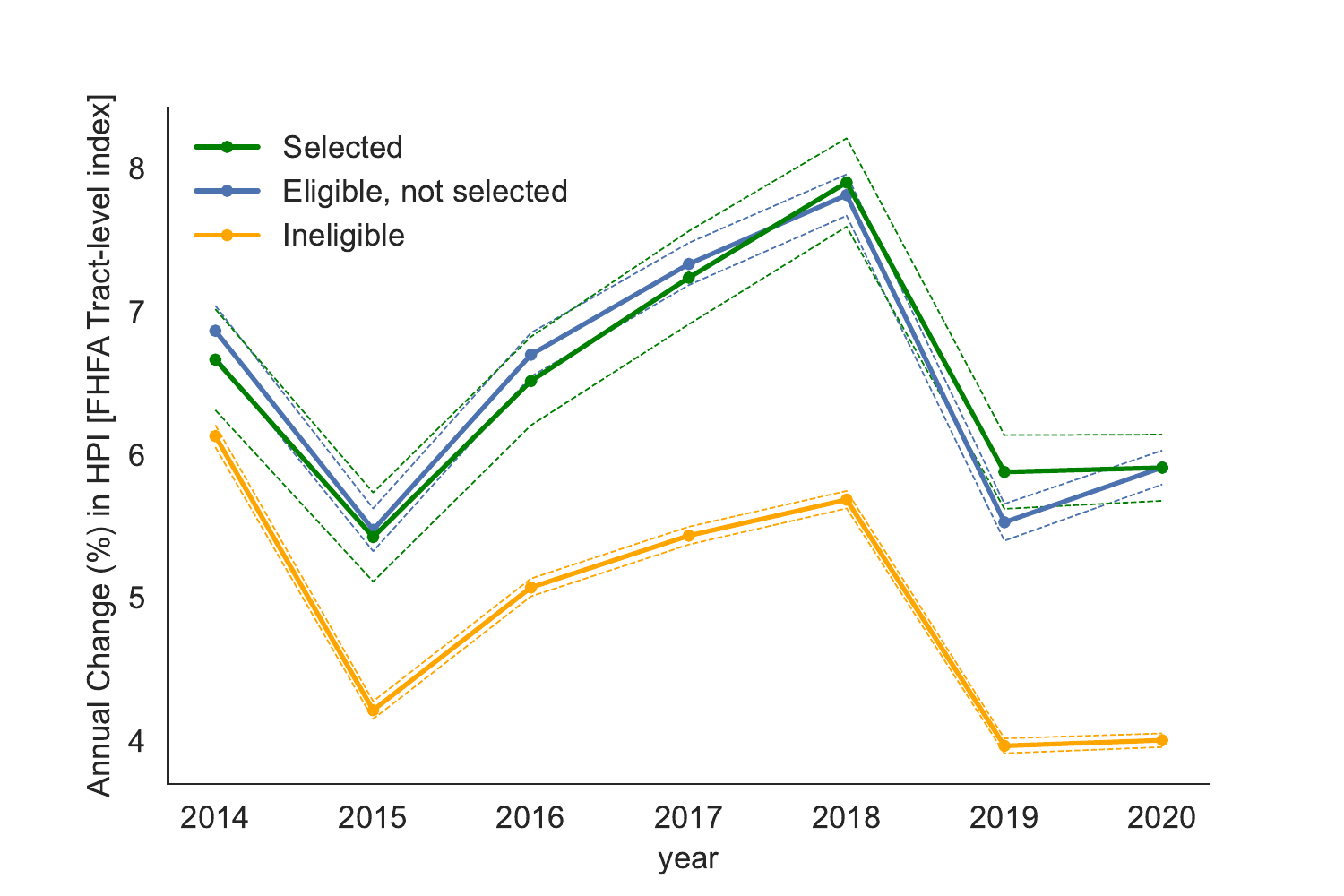}

\vspace{1em}
(b) Event study plot
\includegraphics[width=\textwidth]{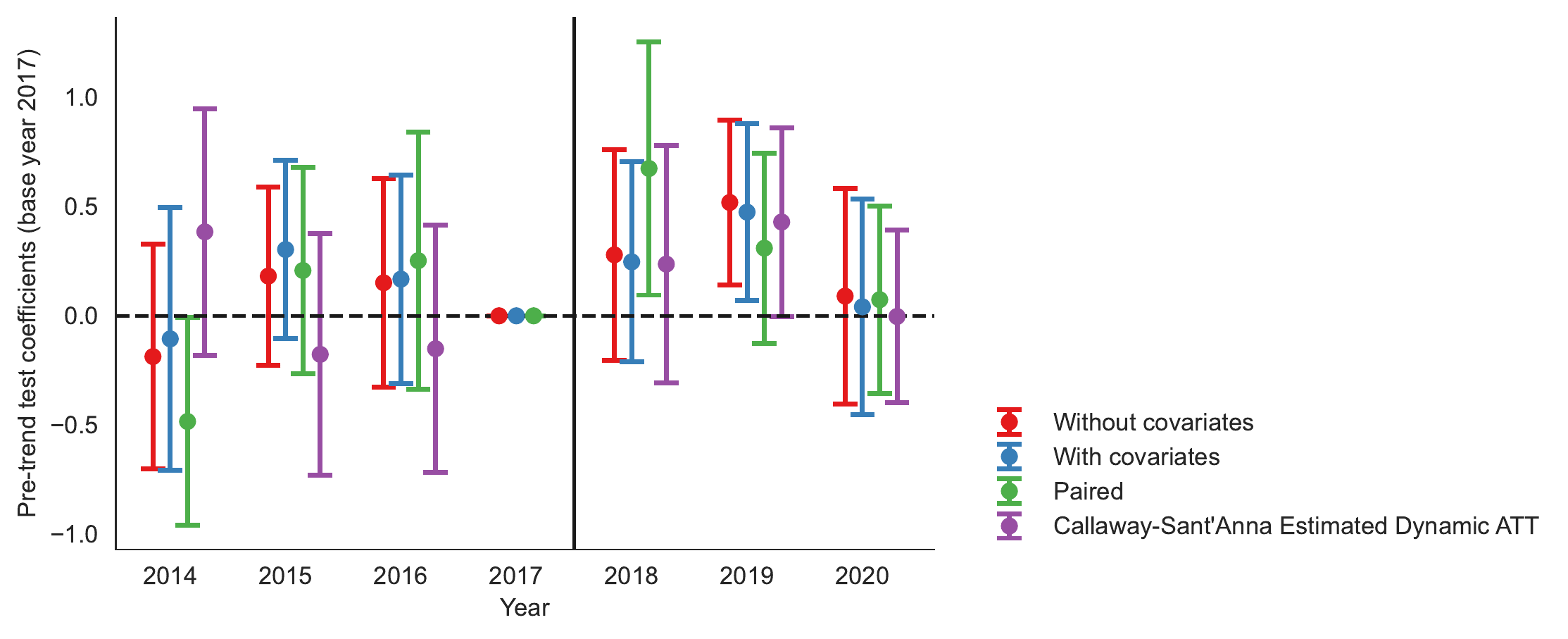}
\caption{Raw trend and event study plot for top panel of  
\cref{tab:tract_and_zip}}
\label{fig:zillow_event_study}
\end{figure}

\afterpage{%
\begin{landscape}

        \begin{table}[tbh]
        \caption{Estimation of ATT using FHFA Tract and ZIP-level data}
        \label{tab:tract_and_zip}
        \scriptsize
        \centering
        \vspace{1em}
        \begin{threeparttable}
        \begin{tabular}{lcccccc}
\toprule
{} &                                     TWFE &                                    TWFE  &                             Weighting CS &                             Weighting DR &                                   Paired &                    Paired (Linear Trend) \\
{} & \hypertarget{tabcol:tract_and_zip1}{(1)} & \hypertarget{tabcol:tract_and_zip2}{(2)} & \hypertarget{tabcol:tract_and_zip3}{(3)} & \hypertarget{tabcol:tract_and_zip4}{(4)} & \hypertarget{tabcol:tract_and_zip5}{(5)} & \hypertarget{tabcol:tract_and_zip6}{(6)} \\
\midrule
\textbf{Tract-level data}        &                                          &                                          &                                          &                                          &                                          &                                          \\
\quad $\hat\tau$                 &              $0.260$ [$-0.034$, $0.553$] &              $0.163$ [$-0.130$, $0.456$] &                $0.212$ $[-0.164, 0.587]$ &                $0.018$ $[-0.374, 0.410]$ &                 $0.359$ $[0.194, 0.523]$ &                $0.285$ $[-0.226, 0.796]$ \\
\quad                            &                                ($0.150$) &                                ($0.144$) &                                ($0.192$) &                                ($0.200$) &                                ($0.084$) &                                ($0.261$) \\
\quad $p$-value                  &                                  $0.089$ &                                  $0.263$ &                                  $0.269$ &                                  $0.929$ &                   $1.900 \times 10^{-5}$ &                                  $0.274$ \\
\quad Pre-trend test $p$-value   &                                  $0.236$ &                                  $0.285$ &                                  $0.495$ &                                      --- &                                  $0.010$ &                                      --- \\
\quad $(N_1, N_0)$               &                            (2917, 10962) &                            (2917, 10962) &                            (2917, 10962) &                            (3095, 11500) &                             (2867, 2867) &                             (2867, 2867) \\
\quad Covariates                 &                                       No &                                      Yes &                                      Yes &                                      Yes &                                      Yes &                                      Yes \\
\quad Sample                     &                    Balanced (2014--2020) &                    Balanced (2014--2020) &                    Balanced (2014--2020) &                    Balanced (2017--2020) &                    Balanced (2014--2020) &                    Balanced (2014--2020) \\
\midrule \textbf{ZIP-level data} &                                          &                                          &                                          &                                          &                                          &                                          \\
\quad $\hat\tau$                 &               $1.181$ [$0.799$, $1.564$] &              $0.342$ [$-0.041$, $0.725$] &                                          &                $0.181$ $[-0.373, 0.735]$ &                                          &                                          \\
\quad                            &                                ($0.195$) &                                ($0.205$) &                                          &                                ($0.283$) &                                          &                                          \\
\quad $p$-value                  &                   $1.497 \times 10^{-9}$ &                                  $0.096$ &                                          &                                  $0.521$ &                                          &                                          \\
\quad Pre-trend test $p$-value   &                                  $0.021$ &                                  $0.653$ &                                          &                                      --- &                                          &                                          \\
\quad $(N_1, N_0)$               &                             (5957, 5758) &                             (5957, 5758) &                                          &                             (1174, 9513) &                                          &                                          \\
\quad Covariates                 &                                       No &                                      Yes &                                          &                                      Yes &                                          &                                          \\
\quad Sample                     &                    Balanced (2014--2020) &                    Balanced (2014--2020) &                                          &                    Balanced (2017--2020) &                                          &                                          \\
\bottomrule
\end{tabular}

        \begin{tablenotes}
        \footnotesize
        \item     \begin{enumerate}

    \item Standard errors are in parenthesis and 95\% confidence intervals are in square brackets.
    Standard errors are clustered at the state level for the tract-level analysis (top panel)
    and clustered at the ZIP level for the ZIP-level analysis (bottom panel).
    Clustering the top panel at the tract level does not qualitatively change results.

    \item Covariates include log median household income, total housing units, percent white,
    percent with post-secondary education,
    percent rental units, percent covered by health insurance among native-born individuals,
    percent below poverty line, percent receiving supplemental income, and percent employed.
    For Column (2), only including log median household income and
percent white as covariates gives $-0.025$ ($0.224$) for the top panel and
$-0.060$ ($0.202$) for the bottom panel.

    \item Pretest for Column (2) interacts covariates with time dummies.

    \item Years 2018 through 2020 are mean-aggregated in Column (4) since the doubly-robust estimation
    only handles two periods.

    \item Discrete treatment in Column (4) is defined as
    the highest 88.3\% of treated
    tract coverage, so as to keep the percentage of treated ZIPs the same as treated tracts.

    \end{enumerate}
    
        \end{tablenotes}
        \end{threeparttable}

        \end{table}
\end{landscape}
}

\section{Results}
\label{sec:res}

The top panel of \cref{tab:tract_and_zip} provides tract-level results, corresponding to
empirical strategies in \cref{sub:twfe,sub:pairs,sub:santanna}. The bottom panel shows
ZIP code level results corresponding to \cref{sub:zip}.  As discussed above, the key
independent variable in the tract level regressions is an indicator variable that takes
on a value of one if the tract is designated an OZ. In the ZIP code level regression,
the key independent variable is the share of the addresses within each ZIP code that lie
within an OZ.

The first two regressions in  \cref{tab:tract_and_zip} show results where the treated
tracts are compared with tracts that were eligible for inclusion within OZs, but were
ultimately not included in the Zones.   These results include time and tract fixed
effects, and the estimated coefficient is 0.26, meaning that prices rose by about
one-fourth of a percentage point annually on average. This coefficient is small in
magnitude, statistically insignificant and precisely enough estimated so that we can
rule out an effect of more than 0.6 percentage points. Table 1 of \cite{cea} performs a
similar analysis, and obtains an estimated coefficient of 0.53 (0.19) for the 2018--2019
data. The size of the discrepancy is within the variation across specifications that we
consider in \cref{tab:tract_and_zip}. The source of the discrepancy is that (i)
\cite{cea} uses the first difference of the FHFA price index, where we use successive
ratios and (ii) we include data up to the year 2020.\footnote{The first difference of
the index is not invariant to the choice of base year. \cite{cea} uses 2013 as the base
year.}

In the second column, we show results allowing for interactions between
tract-level characteristics and year, so that we estimate coefficients on
tract income and other characteristics for each year.    The tract level
covariates do not change over time.   With these added controls, the
coefficient falls to 0.16.  Again, the coefficient is small and
insignificant.   In this case, we can rule out effects of greater than 0.5
percentage points with 95\% confidence.

\Cref{fig:zillow_event_study}(a) shows the tract level housing price
indices visually. The
top two lines show annual growth rates from 2014 to 2020 for the treatment and
control samples that are evaluated in the first two columns of
\cref{tab:tract_and_zip}.  Both of these lines are quite distinct from the third
line, which contains all of the tracts in the U.S. that were never eligible for
OZ status.    The top two lines lie essentially on top of one
another.   Pre-trends appear to be quite similar for the two top groups and
quite different from the third group, which supports the finding of the
pre-trends test reported in \cref{tab:tract_and_zip}.  There is also no visual
change after the law is enacted in 2018.    Both before and after the law is
enacted, price growth in the two groups appears to be almost exactly the same.

The bottom panel of \cref{tab:tract_and_zip} shows results for the ZIP code
analysis.  In this case, the estimated coefficient represents the impact of
moving from having no OZ tracts within the ZIP code to having 100
percent OZ tracts within the ZIP code.\footnote{%
        The average ZIP code has 13.4\% of
        its addresses in a selected Opportunity Zone;
        the median ZIP code has 0.0\%;
        and the 75th percentile has 19.8\%.%
}
The coefficient in the first column is relatively large in the entire table, and it
suggests that as the share of households that live in OZ tracts increase from zero to
one, prices increased by 1.2 percentage points. As the standard errors suggest that the
true coefficient could be as high as 1.5, this coefficient might be economically
meaningful.

Yet there are two reasons to be cautious.  First, the pre-trend test $p$-value
suggests that the parallel trends assumption may be violated.  ZIP codes with a
higher share of addresses residing in OZs seem to diverge from ZIP
codes with a lower share of such households prior to 2018.  Moreover, when we
allow for time varying effects of other tract level characteristics in the
second column, the coefficient becomes 0.34.   That second coefficient is
estimated with enough precision to rule out a coefficient greater than 0.7 at
conventional confidence levels. We interpret these results to suggest that the
ZIP code level analysis also rules out large positive price impacts of
OZ status in 2018.

In the third column, we show the tract-level analyses using the
\cite{callaway2018difference} propensity score weighting method.     The
coefficient estimate is 0.21, and the upper bound on the confidence interval is
0.59.   The fourth regression shows the doubly-robust coefficient estimate that
follows \cite{sant2018doubly}.   The point estimate is 0.02 and the upper bound
estimate is 0.4. The \cite{callaway2018difference} procedure tests for
pre-trends, and we do not reject the  null hypothesis that there is no
pre-trend; the doubly-robust procedure \citep{sant2018doubly} follows a
two-period model, and does not provide a pre-trend test. In both cases, the
results imply that OZ status increased prices by less than one
percentage point in 2018 related to previous years. ZIP-level analysis,
based on discretization of the exposure variable, suggests
effects of similar sizes.

In regressions (5) and (6), we match OZ tracts with the nearest tract that is not in an
OZ.  In some cases, the OZ tracts are matched with the same non-OZ tract. In regression
(5), we find a coefficient of 0.36, which is statistically significant at conventional
levels, but the result is not robust to inclusion of a linear time trend in regression
(6).\footnote{\Copy{lineartrend}{The inclusion of a linear time trend is motivated by
the observation in \cref{fig:zillow_event_study} that the paired design seems to have
nontrivial pre-trend.}} The coefficient falls to 0.28 and the 95\%-confidence interval
rules out a coefficient greater than 0.8.

\Cref{fig:zillow_event_study}(b) shows year-by-year results that correspond to
regressions (1), (2), (3), and (5).   The first two sets of coefficients
show no pre-trend,
but only a small and statistically insignificant increase in price in 2018.  The third
set of coefficients shows a statistically significant price increase in 2018, but
similar-sized fluctuations are present in the pre-treatment period as well (e.g.
2014--2015). Our interpretation of these results is that OZ tracts did experience a
modest increase in price in 2018--2019 relative to the nearest geographic neighbors, but
that this could easily be a reflection of a pre-existing trend or a statistical fluke.

Taken together, these results suggest that if OZ status did generate a positive impact,
that impact was quite small.\footnote{\Copy{representinterp}{Of course,
our results are limited by the FHFA
price coverage at the tract level, which skews towards more populous, richer, and whiter
census tracts (see \cref{asub:represent,asub:unbalanced}). The OZ program may have
differential impacts on places with different characteristics---for instance,
\cite{sage2019opportunity} find differential impact between rural versus urban tracts,
and we find some evidence of differential price impact between residential versus
commercial tracts in \cref{sub:hetero}. Using the ZIP-level data is an attempt at
getting a more representative sample, but it also has limitations due to potential for
spillovers and nonlinear exposure effects.}} There seems to be little possibility that
home buyers anticipated that inclusion in an OZ would have a dramatic impact on the
character of the neighborhood.  This fact does not imply that the OZ program was a
mistake, but rather that it is anticipated to have little effect on the neighborhood.

\subsection{Heterogeneous Impacts of OZ Designation}
\label{sub:hetero}

OZ status confers subsidies to physical investment in a neighborhood. Such subsidies
might have a different impact on housing prices if they largely work by subsidizing
commercial space or if they largely work by subsidizing residential space. If a capital
subsidy increases the presence of job-generating commercial properties, then standard
urban theory predicts that the subsidy will increase residential prices. If the subsidy
increases investment in residential properties, then the impact on housing prices could
be negative.

\afterpage{
        \begin{table}[tbh]
        \caption{Heterogeneous treatment effect by residential population and by housing elasticity}
        \label{tab:hetero_effects_}
        \scriptsize
        \centering
        \vspace{1em}
        \begin{threeparttable}
        \begin{tabular}{lccc}
\toprule
{} &                              No Covariates &                             Few Covariates &                             All Covariates \\
{} & \hypertarget{tabcol:hetero_effects_1}{(1)} & \hypertarget{tabcol:hetero_effects_2}{(2)} & \hypertarget{tabcol:hetero_effects_3}{(3)} \\
\midrule
Treatment $\times$ Post                                 &                 $0.286$ [$0.046$, $0.526$] &                $0.035$ [$-0.204$, $0.275$] &                $0.115$ [$-0.115$, $0.345$] \\
                                                        &                                  ($0.123$) &                                  ($0.122$) &                                  ($0.117$) \\
Treatment $\times$ Post $\times$ Residential            &              $-0.580$ [$-1.024$, $-0.136$] &              $-0.658$ [$-1.101$, $-0.215$] &               $-0.229$ [$-0.643$, $0.185$] \\
                                                        &                                  ($0.227$) &                                  ($0.226$) &                                  ($0.211$) \\
Pretest $p$-value                                       &                                    $0.800$ &                                    $0.269$ &                                    $0.903$ \\
\midrule Treatment $\times$ Post                        &                 $0.388$ [$0.139$, $0.636$] &                $0.131$ [$-0.123$, $0.386$] &                 $0.297$ [$0.048$, $0.547$] \\
                                                        &                                  ($0.127$) &                                  ($0.130$) &                                  ($0.127$) \\
Treatment $\times$ Post $\times$ High Supply Elasticity &              $-0.996$ [$-1.475$, $-0.518$] &              $-1.005$ [$-1.481$, $-0.528$] &              $-0.729$ [$-1.192$, $-0.266$] \\
                                                        &                                  ($0.244$) &                                  ($0.243$) &                                  ($0.236$) \\
Pretest $p$-value                                       &                                    $0.345$ &                                    $0.286$ &                                    $0.663$ \\
\bottomrule
\end{tabular}

        \begin{tablenotes}
        \footnotesize
        \item     \begin{enumerate}

     \item The table reports the regression \[
    Y_{it}^{\obs} = \mu_i +   \alpha_{it} + \tau_0 \one(t\ge t_0, D_i=1) +
    \tau_1
    \one
    (t\ge t_0, D_i=1, R_i = 1) + \gamma  \one(t\ge t_0, R_i = 1)
    \]
    and Treatment $\times$ Post reports $\tau_0$, while Treatment $\times$ Post
    $\times$ Residential reports $\tau_1$. Here $\alpha_{it} = \alpha_t$ in the
    no-covariate specification and $\alpha_{it} = \alpha_{t}'X_i$ in the covariate
    specification. $R_i$ is an indicator for whether the employment to
    residential population ratio is lower than median in the top panel, 
    and is an indicator for whether the price elasticity of housing supply 
    is above median in the bottom panel.

    \item The employment to residential population ratio is
    reported as the ratio of the non-federal employment workforce
    (LEHD-LODES WAC files, 2017, column C000) in
    the tract divided by the population of the tract (ACS 2017 5-year estimates).
    Tracts are classified as residential if the ratio is lower than the median.
    
    \item The price elasticity of housing supply is estimated and provided by 
    \cite{baum2019microgeography}, using housing units.

    \item Standard errors are in parentheses and 95\% confidence intervals are in square brackets.
    Standard errors are clustered at the tract level.

    \item ``All covariates'' consists of log median household income, total housing units, percent white,
    percent with post-secondary education,
    percent rental units, percent covered by health insurance among native-born individuals,
    percent below poverty line, percent receiving supplemental income, and percent employed. ``Few covariates''
    consists of only log median household income and total housing units.
    \end{enumerate}
    
        \end{tablenotes}
        \end{threeparttable}

        \end{table}
        }

Consider a subsidy that decreases the costs of adding residential density, but assume
that existing homes must be bought to provide the land needed to build. If the new
building generates supply but not externalities, then this should decrease the value of
housing units. The value of lower density homes, which make up the bulk of the FHFA
repeat sales properties, could still increase because they are providing land for future
investment. If the new investment generates positive externalities, then there could be
a positive price impact even if supply increases.

We test the hypothesis that OZ status may actually decrease prices by boosting
residential supply by Census tracts in half based on the level of employment to
residential population prior to 2017, where employment is in the LEHD-LODES data. Our
core assumption is that OZ status will act primarily as a subsidy to commercial
properties in the areas that initially have the higher levels of employment to
population and that Zone status will act primarily as a subsidy to residential
construction in the areas where employment to population begins at a lower level.

We test for this heterogeneity in the top panel of \cref{tab:hetero_effects_}. As in the
first two regressions in \cref{tab:tract_and_zip}, the first column includes no
covariates. The second and third columns include fixed covariates that are allowed to
have a different coefficient in each year. In the second column, we control only for log
of median income and total housing units.  In the third column, we include a much wider
set of covariates, and in all cases the estimated coefficients on the covariates are
allowed to vary over time.\footnote{\Copy{covselect}{The choice of covariates in the
second column is more or less arbitrary---it is only meant to show the robustness of the
result, consistent with a similar robustness check in \cref{tab:tract_and_zip}.}} The
top row shows the overall coefficient on OZ status interacted with time, which should be
interpreted as the treatment coefficient for non-residential areas.  The second row
interacts this variable with an indicator variable that takes on a value of one if the
tract is above our sample median in the number of employees per capita in the
residential population.

The first column shows a positive and significant coefficient overall and a negative and
significant interaction term.  These coefficients imply that prices grew by 0.29
percentage points in more commercial OZs, and by $-0.3$ percentage points in more
residential OZs.  The patterns are as expected, but even these coefficients are
relatively modest in magnitude.   In the second column, the positive effect for
commercial tracts disappears when we allow for time varying effects of median income and
the number of housing units.  The negative interaction remains, suggesting the OZ status
had a more negative impact on prices in more residential areas, but again the magnitude
is quite modest.
In the third column, we include a much wider range of controls.  The overall effect is
small in magnitude and statistically insignificant.  The interaction is negative, but
also not statistically significant at standard levels.   The signs are as theory
predicts, but the magnitudes are too small to meaningfully distinguish these effects
from zero.  Moreover, the standards errors are small enough to rule out truly large
effects in either direction.

\Copy{baumsnow}{Similarly, the differential impact of the OZ designation should be more
pronounced if we segment the sample by housing supply elasticity. Intuitively, areas
with high supply elasticity respond to the changing incentives by building more housing,
thereby we should expect housing prices to decrease, relative to areas with low supply
elasticity. This is indeed what we observe in the bottom panel of
\cref{tab:hetero_effects_}: Across all three configurations of covariates, the price
treatment effect for high elasticity tracts (per the estimates by
\cite{baum2019microgeography}) tends to be smaller than that for low elasticity tracts.
The estimates continue to  rule out large positive price effects, even for census tracts
with low elasticity of supply.}

Once again, there seems to be little evidence to support the view that
OZ status generated the expectation, at least among home buyers,
that these areas would transition from poverty to prosperity.

\subsection{Effect on residential permitting} 
\label{sub:permittingresults}

\afterpage{%
\begin{landscape}
    \begin{figure}[tb]
    \centering
    \includegraphics[width=1.3\textheight]
    {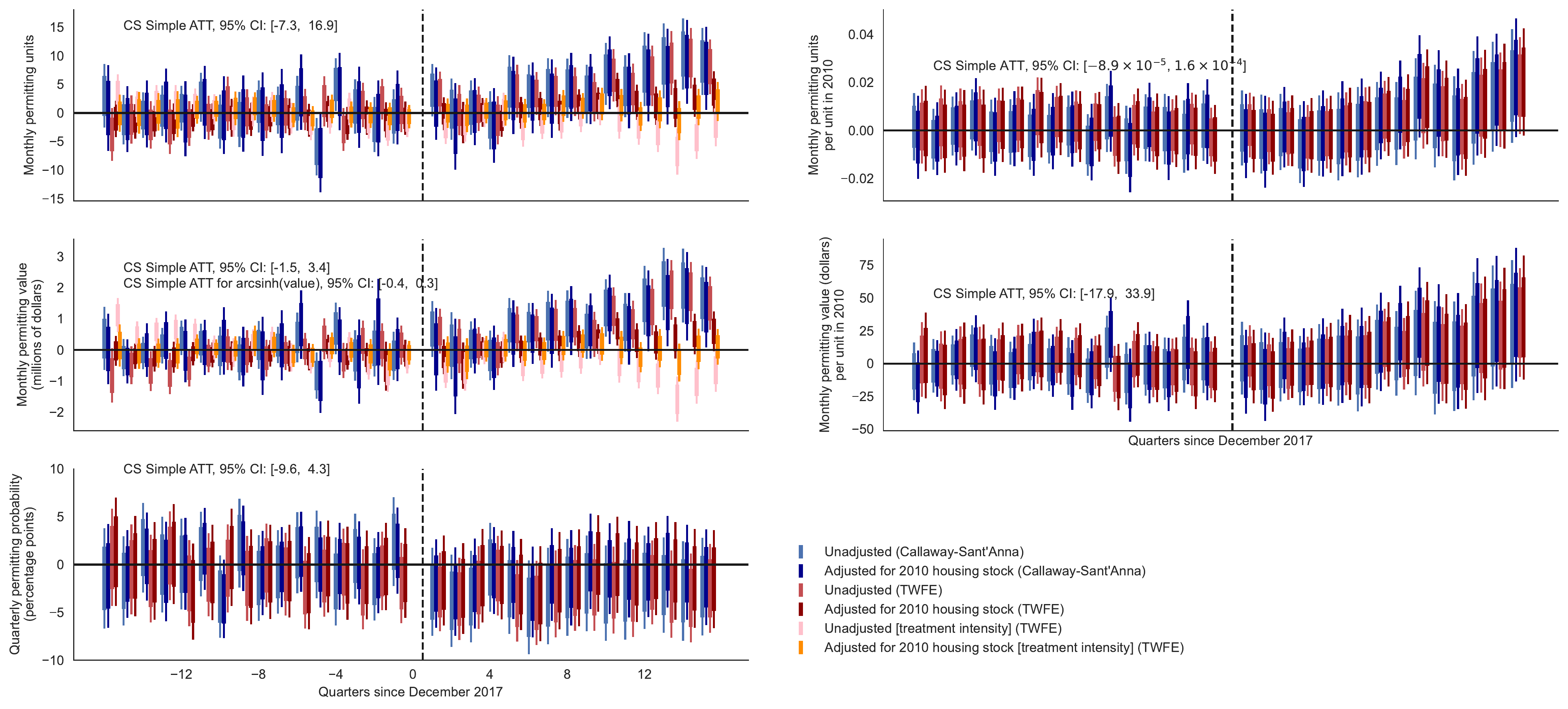}
    
    \begin{proof}[Notes]
        Thick error bars are conventional pointwise (1.96 $\times$ SE) Wald
    intervals, whereas the thin error bars are simultaneous 
    (max-$t$) confidence bands
    \citep{montiel2019simultaneous}.
    \end{proof}

    \caption{Event study plot as well as ATT estimates for effect on
    various measures of
    residential permitting}
    \label{fig:quant}
\end{figure}    

\end{landscape}
}

Another dimension of OZs' effect on the housing market is through
its effect on quantities, which we measure via housing permits. Large increases
in permitting---and hence increase in housing supply---may explain the lack of
price impact that we observe. New housing may indeed be evidence that the
neighborhood is changing, although without price effects it is hard to new if
the neighborhood is changing for the better.   Presumably, part of the point of
a capital subsidy is to induce new capital formation and housing permits provide
our best indication of whether the OZ subsidy is generating new
capital.

Working with housing permits brings a few econometric challenges. Permits can either be
measured in terms of units of housing or value of housing. Permitting is extremely
sparse---more than half of place-by-month cells have zero permitting---and
heterogeneous, as certain populous places have much more permitting than others. These
challenges result in an ambiguity of the right response variable to use, and parallel
trends assumptions on different response variables may not be internally consistent with
each other \citep{roth2020parallel}. These concerns notwithstanding, we present
event-study plots for a range of outcomes in \cref{fig:quant}. Each subplot of
\cref{fig:quant} is an event-study plot for a particular response variable showing the
estimated coefficients for a given three-month period. For each response variable, we
consider two specification (with and without adjusting for housing stock in 2010) and
two estimators (\cite{callaway2018difference} and two-way fixed-effect). Qualitatively
speaking, the four settings have similar estimated coefficients.

The left column of \cref{fig:quant} consists of event study plots for outcomes that are
not normalized by city size. The first two rows of the left column are effects for
monthly permitting units and value in levels. The results are consistent across the two
measures of permitting volume, and point towards a small positive effect, relative to
the last untreated period, of 5 units or \$900,000 in value per month, though it may be
as low as 1 unit or \$200,000 in value.\footnote{These estimates assume perfect parallel
trends, which may be substantially attenuated if we consider effects relative to
plausible pretrends \`a la \cite{roth2019honest}.} Though, notably, these
results are not very robust to adjusting for size of housing stock or to
using a continuous exposure variable for treatment definition. In
particular, many point estimates are negative when treatment is defined as
the exposure to OZs.

This does suggest that OZs may have increased housing supply in treated areas,  but
unusual statistical features like sparse and long-tailed data make these results
somewhat unreliable.  As robustness checks we consider a range of measures that are
better behaved statistically. First, we consider transforming value via the
$\mathrm{arcsinh}$ function, which is an ad hoc approximation of the logarithm that is
defined at zero. Doing so results in a confidence interval of $[-0.4, 0.3]$, which means
that we cannot rule out zero, and our point estimate is negative.

\Copy{whynormalize}{On the right side of \cref{fig:quant}, we consider
units and value normalized by a measure of city size (number of housing units in 2010),
 motivated by concerns that the effect in pure quantity terms may be driven by large
 Census places.} In this case, we cannot reject zero impact. A separate analysis for
 quantile treatment effects with changes-in-changes
 \citep{athey2006identification,callaway2016quantile} in \cref{fig:cic} in
 \cref{asec:cic} similarly finds positive but small quantile effects.

Finally, in the bottom panel of the left column of \cref{fig:quant}, we consider
the response variable of non-zero permitting in a given three-month period, and
we find a mildly negative point estimate of the ATT relative to the last
untreated period, but it is clear that this effect is small relative to the
uncertainty from the event-study plot.

\Copy{persistence}{As an aside, viewing the price and permitting results
together, it also seems that the permitting effects---if they do exist---may be more
persistent than the price effects. This is consistent with the fact that prices should
adjust quickly, reflecting expectations, and so effects on price \emph{growth} are
naturally transitory. Quantities, on the other hand, adjust more slowly to changes in
neighborhood policy. }

We conclude that there is weak evidence that OZ status
increased residential permitting.
The basic effects suggest that OZs did encourage permitting, but the results were not
robust enough for us to have much confidence.  Moreover, since the results disappear when we looked
at permits normalized for the 2010 stock, the effects are probably driven by a few cases
of quite large bursts of permitting in relatively small places.

\section{Cost-benefit analysis}
\label{sec:costbenefit}

We have too little information to fully assess the OZ program, but we can
compare the program's cost in foregone taxes with its impact on property values.  Land
value provides the most canonical welfare measure in an urban setting.  \cite
{arnott1979aggregate} show that land value captures social welfare in a small open city
when people are identical or when there are no infra-marginal residents.   These
assumptions are unlikely to be met in reality, and our parameter estimates do not allow
us to fully capture the impact of the OZ program on land values. OZ designation creates
incentives for business-related investment and the employees, owners and customers of
these businesses may well live outside of the OZ.    Yet our estimates
look only at the impact of the OZ program on property within the zone itself, and we
look only at the impact of the program on residential housing, not on unoccupied or
industrial land.   

Nonetheless, we will now do two simple calculations estimating the total increase in
housing wealth in OZ tract based on our estimates, and the value created
by new housing construction. The increase in housing wealth will reflect both local
amenities created by the program and the tax benefits reaped by home buyers who are
participating in the OZ program.   We will estimate benefits on a per unit basis and
then scale those benefits up by the number of units.  As we have discussed, 8,764
Census tracts were allocated to OZs.  These tracts contained a total of
12.78 million housing units.  

We estimated increases in housing price as being proportional the to the value of those
units.    The average value of owner-occupied housing in OZ tracts
is \$155,000, but only 44 percent of OZ units are owner-occupied.   We do not have
direct measures of the value of rental properties and those properties are likely to be
less valuable than owner-occupied units.  To address this issue, we ran a housing price
hedonic using owner-occupied units from American Community Survey during the
pre-period.\footnote{Specifically, at the tract level, we regress median owner-occupied housing price on average household size, number of rooms, median year built, and median income of owner-occupied housing units. We then use the estimated regression coefficients to predict median renter-occupied housing price. Such a regression yields a reasonable approximation of value of rental units if the conditional relationships between median price and housing characteristics are similar across renter- and owner-occupied units.} We then used the estimated coefficients to price the rental units in
OZs.  Our predicted value for rental units is 60 percent of the predicted
value relative to owner-occupied units.  Consequently, we will assume that the average
pre-treatment value of rental units equals \$90,000 and that the treatment effect of
OZ status as a percentage of value was the same for owner and rented
properties.  Taken together, these estimates suggest that the total housing value in
OZs during the pre-period was \$1.5 trillion.   

\begin{figure}
    \centering
    \includegraphics[width=\textwidth]{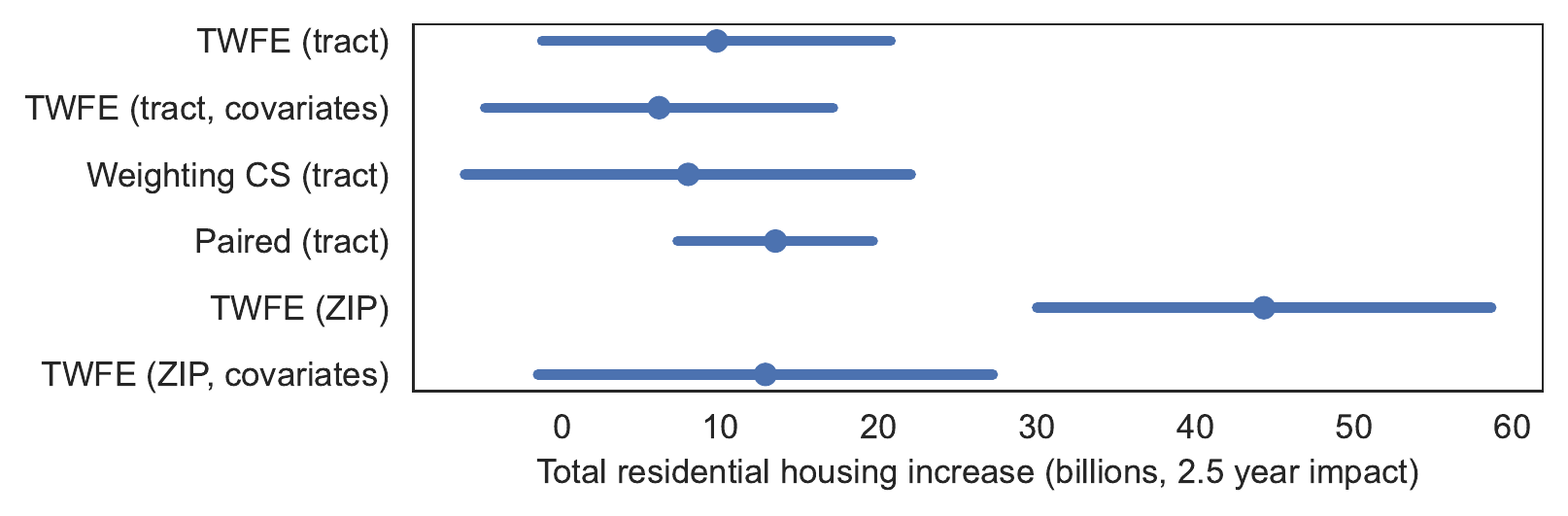}
    \caption{Estimates of value impact on existing units}
    \label{fig:qdelp}
\end{figure}

\Copy{deltahousevalue}{Our different point estimates then suggest total increases in total residential housing
value that are summarized in \cref{fig:qdelp}. The preferred point estimates yield value
increases that range from \$6 to \$15 billion, with \$44 billion being a clear outlier.
The upper bound point estimates provide value creation between \$17--27 billion, with
\$60 billion being a clear outlier. }

\Copy{harberger}{
We can augment these values by considering the value of new housing created according to
our permit data. We use the treatment effects per unit of housing stock estimates in
\cref{fig:quant} and compute that the total value of new housing is \$0.5--1 billion,
with upper bounds that range from \$2.5--4 billion. This value created is not a welfare
calculation, but it does reflect
a different way of calculating the effect of the housing. 

To convert this new construction into a welfare calculation, we would need to go from
this total output number to a measure of consumer welfare and producer profit and would
need to convert this annual flow into a sum of the life of the project.   While industry
rule of thumb calculations typically suggest a profit margin of 20 percent, that margin
is better seen as reflecting accounting profits rather than true economic profits.    
If we took 1/5 of total production as reflecting total social welfare created by this
new production, then this would imply annual profit flows ranging from \$200 million to
\$800 million dollars.

This figure is much larger than the classic welfare triangle suggested by this
calculation.  The change in the number of units would range from 4,500 to 20,000 from
our lower preferred estimate to our highest upper bound estimate.  The largest change
in price per unit point estimate is \$4,500 over 2.5 years.  By multiplying the change
in price by the change in units by $1/2$, we find a total ``Harberger'' triangle
of \$10 to \$45 million. This is fairly negligible relative to the other figures, and
so we will ignore it going forward. }

So far, we have calculated the net price impact on existing quantities, $q\Delta p
\approx \$10 \text{ billion}$. This
corresponds to Marshallian surplus with perfectly inelastic housing supply. In general,
the Marshallian surplus is, to a first-order, $W = \pr{1 + |\epsilon_S/\epsilon_D|}
q \Delta p $, where $\epsilon_S, \epsilon_D$ are supply and demand elasticities,
respectively. Point estimates for changes in quantity and price suggest the supply
elasticity is anywhere between 0.03 to 0.2, which is sensitive to estimation of the
price impact. \cite{baum2019microgeography} suggest the supply elasticity is about 0.4.
On the other hand, modelling the OZ program as a demand shock implies that the demand
elasticity is not identified from our data, but estimates by \cite
{hanushek1980price} and \cite{albouy2016housing} suggest $\epsilon_D$ values ranging
from $-0.4$ to $-0.6$. Thus, $W \le 2q\Delta p$ seems like a reasonably generous upper
bound, which puts the consensus point estimate at about \$20 billion, with consensus
upper bounds going as high as \$60 billion. 

\Copy{caution}{We caution that the benefit analysis above err on the side of optimism,
and there are
reasons to think they overstate the positive impact of the OZ program. First, the
statistical
uncertainty in our price estimates cannot reject a welfare impact of zero across many
specifications. Second, as the identification strategy compares OZs to similar,
untreated tracts, we cannot rule out the possibility that the price impact of OZs comes
from displacement of investment that would have gone to the Census tracts that we use
as controls. If investors substitute from these control tracts to OZs, welfare impacts
from our estimates overstate the national net value creation of the program. Third,
this analysis ignores heterogeneity, and there is evidence from 
\cref{tab:hetero_effects_} that the price impact for residential tracts---where existing
residential units concentrate---may be smaller than the average price impact or even be
negative. This means that the housing-unit level price impact may be lower than the
tract-level price impact in \cref{tab:tract_and_zip} that we use for the benefit
analysis.}

 \subsection{The tax expenditure costs of the program}

A report from the \cite{cea} ``estimates that the Federal Government forgoes \$0.15
for every
\$1 in capital gains invested in a Qualified Opportunity Fund before 2020, or about
\$11.2 billion for the \$75 billion raised through the end of 2019.''    \cite{arefeva2021job} suggest
that this number should be increased five-fold because Qualified Opportunity Funds are responsible for only twenty percent of the investment in Opportunity Zones.  Yet the Council of Economic Advisors estimate of \$75 billion may also radically overstate the true level of Opportunity Zone investment, because it is based on scaling upward actually observed Qualified Opportunity Fund levels of \$7.6 billion and \$2.9 billion observed by Novogradac, an accounting firm, and the Securities and Exchange Commission respectively.
\cite{kennedy2021neighborhood} report only \$18.9 billion in total OZ investments on tax records that were filed
electronically by businesses in 2019.  That report suggests that this represents 75
percent of all OZ investments, which would imply \$25 billion in investment.     If this
\$25 billion represents the total stock of investment than the total cost would be less
 than \$4 billion.  If that figure represents the annual flow of new investment in OZs,
 then the annual cost would come close to the \$3.75 billion, which is reasonably close
 to a Joint Committee on Taxation figure of \$3.5 billion per year cited in \cite{kennedy2021neighborhood}.  
 Depending on the discount factor, \$3.5 billion per year could be
 significantly higher than the \$55 billion cited by \cite{arefeva2021job}.  If the
 costs of this program are more than \$50 billion, then this seems significantly higher
 than most estimates of increase in housing values.  The welfare associated with new
 construction is
 more difficult to assess, but it can be compared with the \$3.5 billion flow
 discussion by \cite{kennedy2021neighborhood}.

\section{Conclusion}
\label{sec:conc}

OZs are America's most important new national spatial policy since the Empowerment Zone
program began during the Clinton era. They are intended to spur investment in high
poverty areas. The hope of this program is that it would generate neighborhood revival,
yet we find little evidence to support this view at this early date. Housing prices may
have gone up in OZ areas after their enactment in 2018, but if they did the overall
price impact seems to have been less than one percentage point. We find suggestive
evidence that OZ status increased prices in more commercial areas and reduced prices in
more residential areas, presumably because Zone status generated a subsidy for building
new homes. We also found some evidence suggesting that OZs did encourage
new building.
Of course, even if OZs did induce new building in treated areas, it is not obvious if
that outcome is socially desirable.

The designation of the OZ tracts was only made public in the
summer of 2018. Consequently, our results reflect 27 months of subsequent data,
during which the world experiences a once-in-a-century global pandemic. These
features of our results should make us cautious about any interpretation. We are
at an early point and home price effects can, at best, capture the expectations
about neighborhood change held by recent home buyers. These buyers could be
wrong: In the future OZ status could end up correlated with
neighborhood upgrading. 

\Copy{welfareconc}{Our preliminary estimates suggest that a generous welfare estimate of OZ's impact over
2.5 years is \$20 billion, versus about \$3.75 billion of cost per annum estimated by 
\cite{kennedy2021neighborhood}---aggregating to about \$10 billion over 2.5 years. Under
 such an optimistic scenario, it does seem that the OZ program is beneficial. However,
 the uncertainty in these welfare means that we cannot reject a welfare impact of zero.
 There is also considerable uncertainty in the cost estimates: Our welfare estimates
 are generally comparable to the low-end of cost estimates, but are much less than the
 high end. Moreover, much of the estimated welfare increase comes from increasing value
 of the housing stock, but we may not expect the price effect to persist if it mainly
 reflects buyers' expectations, whereas the costs of the program continue to accrue.}

\newpage

\vfill

\newpage

\bibliographystyle{chicago}
\bibliography{refs}

\newpage

\appendix

\section{Covariates, summary statistics, and balance}

\begin{figure}[tb]
\centering
\includegraphics[width=1\textwidth]{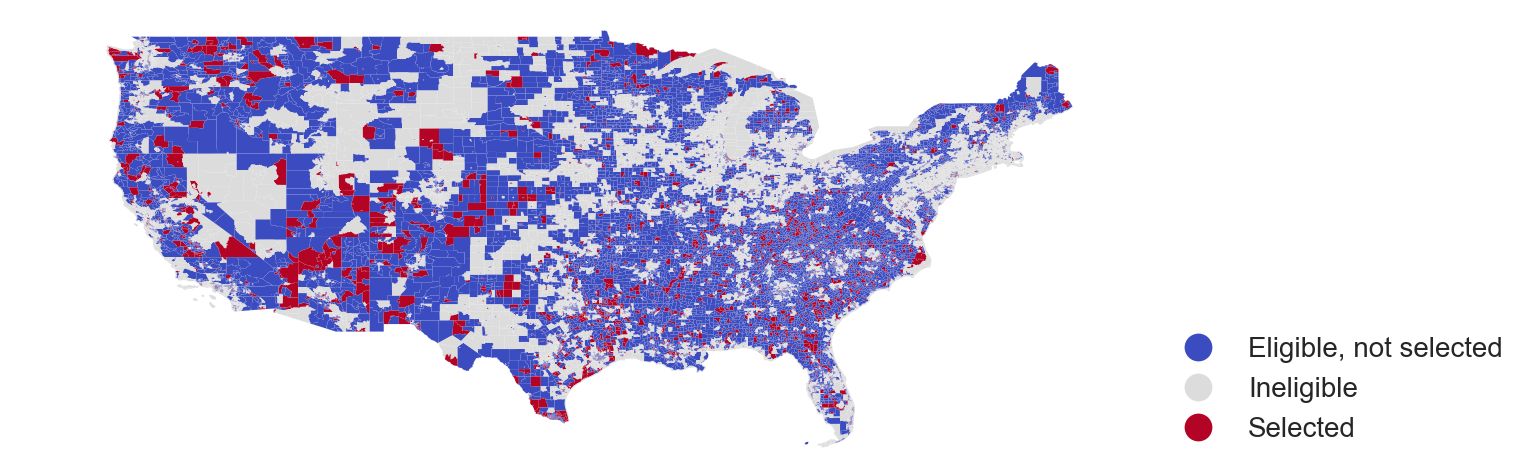}
\caption{Map of Opportunity Zones status}
\label{fig:map}
\end{figure}

\subsection{Variable definitions}
Variable definitions for covariates and simple calculations, along with their associated
code used in the Census API, are shown in \cref{tab:defs}. Summary statistics and
covariate balance are shown in
\cref{tab:balance_of_selected_opportunity_zones_and_eligible_census_tracts,tab:paired_balance}.
We see that compared to the control group, the selected OZs are less populated, less
employed, less likely to attain higher education, have more rental units, and are less
wealthy. Similar trends persist when compared to their non-selected geographical
neighbors in \cref{tab:paired_balance}. While the covariate non-balance threatens
identification by making the (conditional) parallel trends assumption less plausible, it
does not reject parallel trends either. Since identification requires only trends to be
parallel and allows for level differences, it also allows for level differences in
observed or unobserved characteristics, so long as the trends are the same
(conditionally).

\subsection{Estimation with unbalanced panel}
\label{asub:unbalanced}

\Copy{unbalanced}{
The counterpart to the first two columns of \cref{tab:tract_and_zip} with
unbalanced panel is shown in \cref{tab:unbalanced_}, which does not
qualitatively change conclusions.}

        \begin{table}[tbh]
        \caption{TWFE results with unbalanced panel of tract-level data}
        \label{tab:unbalanced_}
        
        \centering
        \vspace{1em}
        \begin{threeparttable}
        \begin{tabular}{lcc}
\toprule
{} &                                   TWFE &                                  TWFE  \\
{} & \hypertarget{tabcol:unbalanced_1}{(1)} & \hypertarget{tabcol:unbalanced_2}{(2)} \\
\midrule
$\hat\tau$               &            $0.219$ [$-0.062$, $0.499$] &            $0.099$ [$-0.181$, $0.380$] \\
                         &                              ($0.143$) &                              ($0.147$) \\
$p$-value                &                                $0.133$ &                                $0.502$ \\
Pre-trend test $p$-value &                                $0.980$ &                                $0.746$ \\
$(N_1, N_0)$             &                          (3806, 13983) &                          (3806, 13983) \\
Covariates               &                                     No &                                    Yes \\
Sample                   &                Unbalanced (2014--2020) &                Unbalanced (2014--2020) \\
\bottomrule
\end{tabular}

        \begin{tablenotes}
        \footnotesize
        \item     \begin{enumerate}

    \item Standard errors are in parenthesis and 95\% confidence intervals are in square brackets.
    Standard errors are clustered at the state level for the tract-level analysis (top panel).
    Clustering the top panel at the tract level does not qualitatively change results.

    \item Covariates include log median household income, total housing units, percent white,
    percent with post-secondary education,
    percent rental units, percent covered by health insurance among native-born individuals,
    percent below poverty line, percent receiving supplemental income, and percent employed.

    \item Pretest for Column (2) interacts covariates with time dummies.
    \end{enumerate}
    
        \end{tablenotes}
        \end{threeparttable}

        \end{table}

\subsection{Representativeness of the tract sample}
\label{asub:represent}

\begin{figure}[tb]
    \centering
    \includegraphics{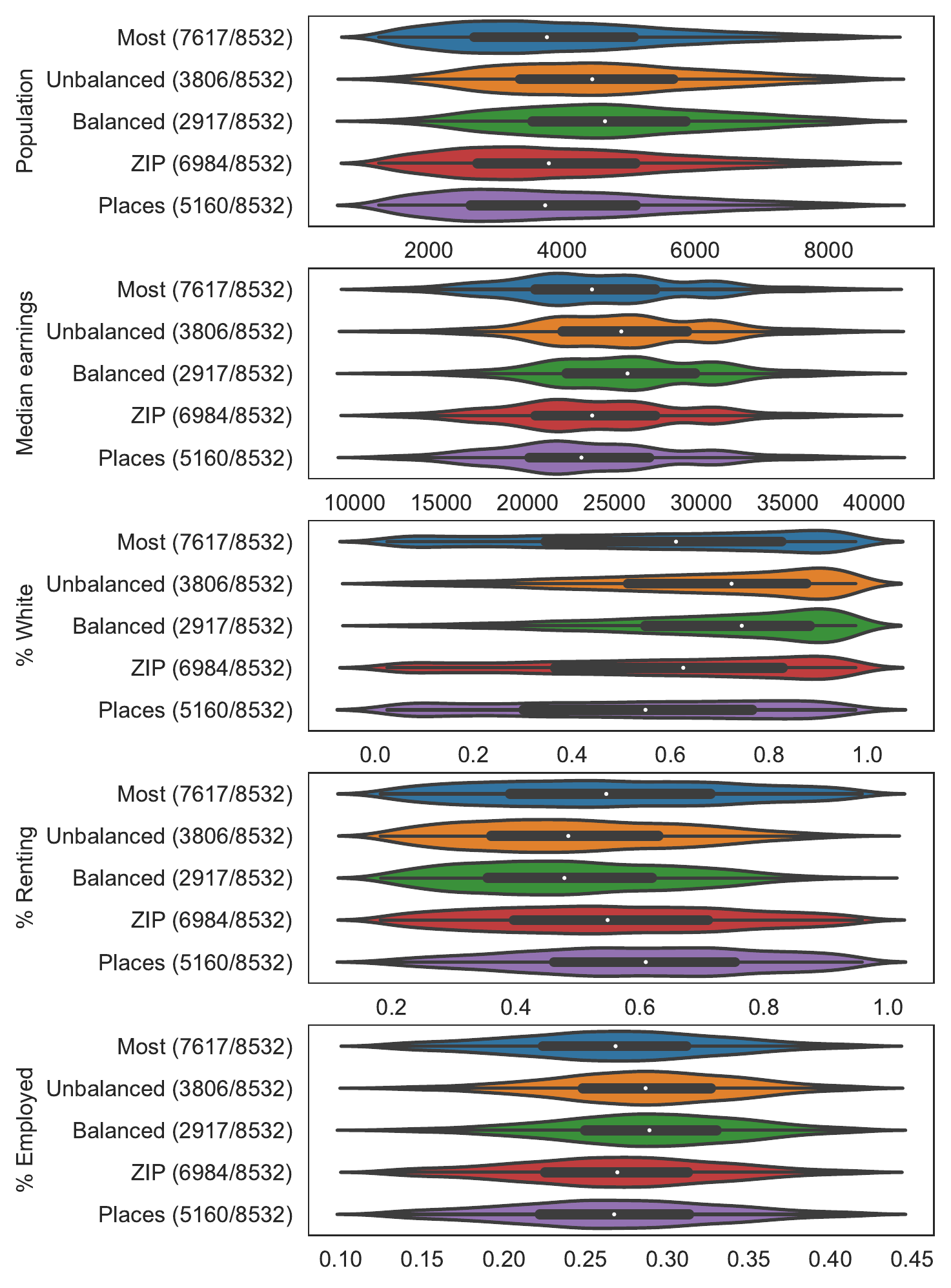}
    \caption{Distribution of covariates in different subsamples of the
    tract data (Middle 95\% of each covariate shown)}
    \label{fig:tract_covariate_diff}
\end{figure}

\begin{figure}[tb]
    \centering
    \includegraphics{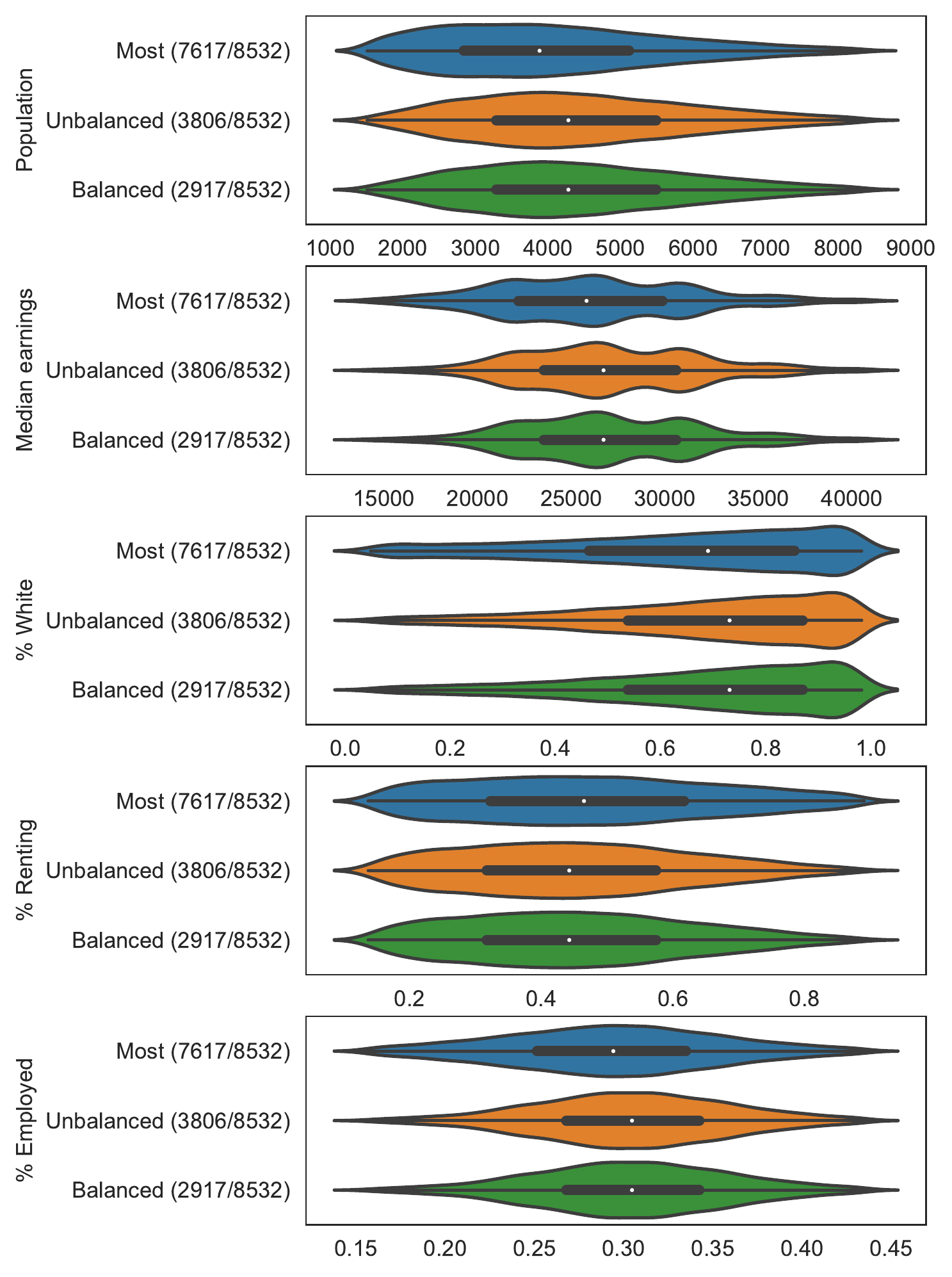}
    \caption{Distribution of covariates in different subsamples of the
    \emph{untreated} tract data (Middle 95\% of each covariate shown)}
    \label{fig:tract_covariate_diff_control}
\end{figure}

\Copy{represent}{
We plot the distribution of a few covariates in the overall sample of selected
OZs\footnote{Among which 7,617 have ACS covariate data coverage.} versus various
subsamples with price data in \cref{fig:tract_covariate_diff}; we show the covariate
means in \cref{tab:cov_diff}. We see that, roughly speaking, \begin{itemize}
    \item Treated tracts with price data tend to be more populous (by about
    25\% on average), richer (by 8\% / \$2,000 in median earnings), whiter (by 12
    percentage points), less renter-occupied (by 7 percentage points), and more employed
    (by 3 percentage points).
    \item The distributions of the covariates in the subsample with
    price data still has substantial overlap with the overall distribution
    (\cref{fig:tract_covariate_diff}), as opposed to the distribution in the subsample
    being some truncation of the overall distribution.
    \item Tracts in the unbalanced panel versus the subsample of tracts in
    the balanced panel look quite similar. 
    \item Tracts covered by the ZIP code level data in 2018 (we
    have covariate data for 6984 out of 6988) look quite similar to the
    overall sample of selected OZ tracts. 
    \item Tracts covered by the Census place level data (we have covariate
    data for 5160 out of 5162) look likewise similar to the overall sample,
    perhaps somewhat less so than the ZIP-level data. It tends to skew less
    white and more rent-occupied than the overall population.
\end{itemize}
Similarly, we show the covariate distribution discrepancies in the control
group (eligible, but not selected census tracts) in 
\cref{fig:tract_covariate_diff_control}. The qualitative patterns are
similar to that of the treated group.
}

\afterpage{\begin{landscape}
    \begin{table}[tbh]
        \caption{Covariate means by subsample in the tract data (Covariate data from ACS 2017)}
        \label{tab:cov_diff}
        \centering
        \vspace{1em}
        \begin{tabular}{lcccccc}
\toprule
{} &                   \texttt{level\_1} &           \texttt{Most (7617/8532)} &     \texttt{Unbalanced (3806/8532)} &       \texttt{Balanced (2917/8532)} &            \texttt{ZIP (6984/8532)} &         \texttt{Places (5160/8532)} \\
{} & \hypertarget{tabcol:cov_diff1}{(1)} & \hypertarget{tabcol:cov_diff2}{(2)} & \hypertarget{tabcol:cov_diff3}{(3)} & \hypertarget{tabcol:cov_diff4}{(4)} & \hypertarget{tabcol:cov_diff5}{(5)} & \hypertarget{tabcol:cov_diff6}{(6)} \\
\midrule
\texttt{population}       &                                mean &                          $4018.494$ &                          $4743.423$ &                          $4933.514$ &                          $4055.862$ &                          $4022.165$ \\
\texttt{population}       &                                 std &                          $1970.423$ &                          $1889.225$ &                          $1900.284$ &                          $1940.016$ &                          $2053.094$ \\
\texttt{population}       &                               count &                             $7,617$ &                             $3,806$ &                             $2,917$ &                             $6,984$ &                             $5,160$ \\
\texttt{population}       &                                  SE &                            $22.582$ &                            $30.623$ &                            $35.184$ &                            $23.217$ &                            $28.587$ \\
\texttt{median\_earnings} &                                mean &                         $24150.852$ &                         $25629.700$ &                         $26037.611$ &                         $24081.051$ &                         $23873.037$ \\
\texttt{median\_earnings} &                                 std &                          $7076.359$ &                          $6203.688$ &                          $6253.787$ &                          $6952.603$ &                          $7337.118$ \\
\texttt{median\_earnings} &                               count &                             $7,617$ &                             $3,806$ &                             $2,917$ &                             $6,984$ &                             $5,160$ \\
\texttt{median\_earnings} &                                  SE &                            $81.187$ &                           $100.558$ &                           $115.791$ &                            $83.290$ &                           $102.270$ \\
\texttt{pct\_white}       &                                mean &                             $0.568$ &                             $0.676$ &                             $0.699$ &                             $0.577$ &                             $0.510$ \\
\texttt{pct\_white}       &                                 std &                             $0.299$ &                             $0.246$ &                             $0.230$ &                             $0.294$ &                             $0.292$ \\
\texttt{pct\_white}       &                               count &                             $7,617$ &                             $3,806$ &                             $2,917$ &                             $6,984$ &                             $5,160$ \\
\texttt{pct\_white}       &                                  SE &                             $0.003$ &                             $0.004$ &                             $0.004$ &                             $0.004$ &                             $0.004$ \\
\texttt{pct\_rent}        &                                mean &                             $0.557$ &                             $0.488$ &                             $0.479$ &                             $0.555$ &                             $0.615$ \\
\texttt{pct\_rent}        &                                 std &                             $0.222$ &                             $0.181$ &                             $0.176$ &                             $0.215$ &                             $0.208$ \\
\texttt{pct\_rent}        &                               count &                             $7,617$ &                             $3,806$ &                             $2,917$ &                             $6,984$ &                             $5,160$ \\
\texttt{pct\_rent}        &                                  SE &                             $0.003$ &                             $0.003$ &                             $0.003$ &                             $0.003$ &                             $0.003$ \\
\texttt{pct\_employed}    &                                mean &                             $0.266$ &                             $0.289$ &                             $0.292$ &                             $0.267$ &                             $0.267$ \\
\texttt{pct\_employed}    &                                 std &                             $0.077$ &                             $0.064$ &                             $0.065$ &                             $0.076$ &                             $0.080$ \\
\texttt{pct\_employed}    &                               count &                             $7,617$ &                             $3,806$ &                             $2,917$ &                             $6,984$ &                             $5,160$ \\
\texttt{pct\_employed}    &                                  SE &              $8.789 \times 10^{-4}$ &                             $0.001$ &                             $0.001$ &              $9.072 \times 10^{-4}$ &                             $0.001$ \\
\bottomrule
\end{tabular}

        \end{table}

    \begin{table}[tb]
    \scriptsize
    \caption{Data sources}
    \label{tab:data_sources}
    \centering
\vspace{1em}
    \begin{tabular}{cc}
    \toprule
    Data Source  & URL (Accessed 2022-03-18) \\ \midrule
FHFA Housing Price Index & \url{https://www.fhfa.gov/DataTools/Downloads/Pages/House-Price-Index-Datasets.aspx} \\ 
Urban Institute OZ Data & \url{https://www.urban.org/policy-centers/metropolitan-housing-and-communities-policy-center/projects/opportunity-zones} \\ 
Longitudinal Employer-Household Dynamics & \url{https://lehd.ces.census.gov/data/} \\ 
& \url{https://datacatalog.urban.org/dataset/longitudinal-employer-household-dynamics-origin-destination-employment-statistics-lodes} \\
TIGER Geographic Shapefiles & \url{https://www.census.gov/geographies/mapping-files/time-series/geo/tiger-line-file.html} \\ 
Census Business Patterns & \url{https://www.census.gov/construction/bps/} \\ 
Building Permits Survey & \url{https://www.census.gov/construction/bps/} \\
Missouri Census Data Center, Geocorr 2000 & \url{https://mcdc.missouri.edu/applications/geocorr2000.html} \\ \bottomrule
    \end{tabular}
\end{table}
\end{landscape}
}
\section{Including all OZs}
\label{asec:all_oz}

\begin{figure}[tb]
    \centering
    \includegraphics[width=\textwidth]{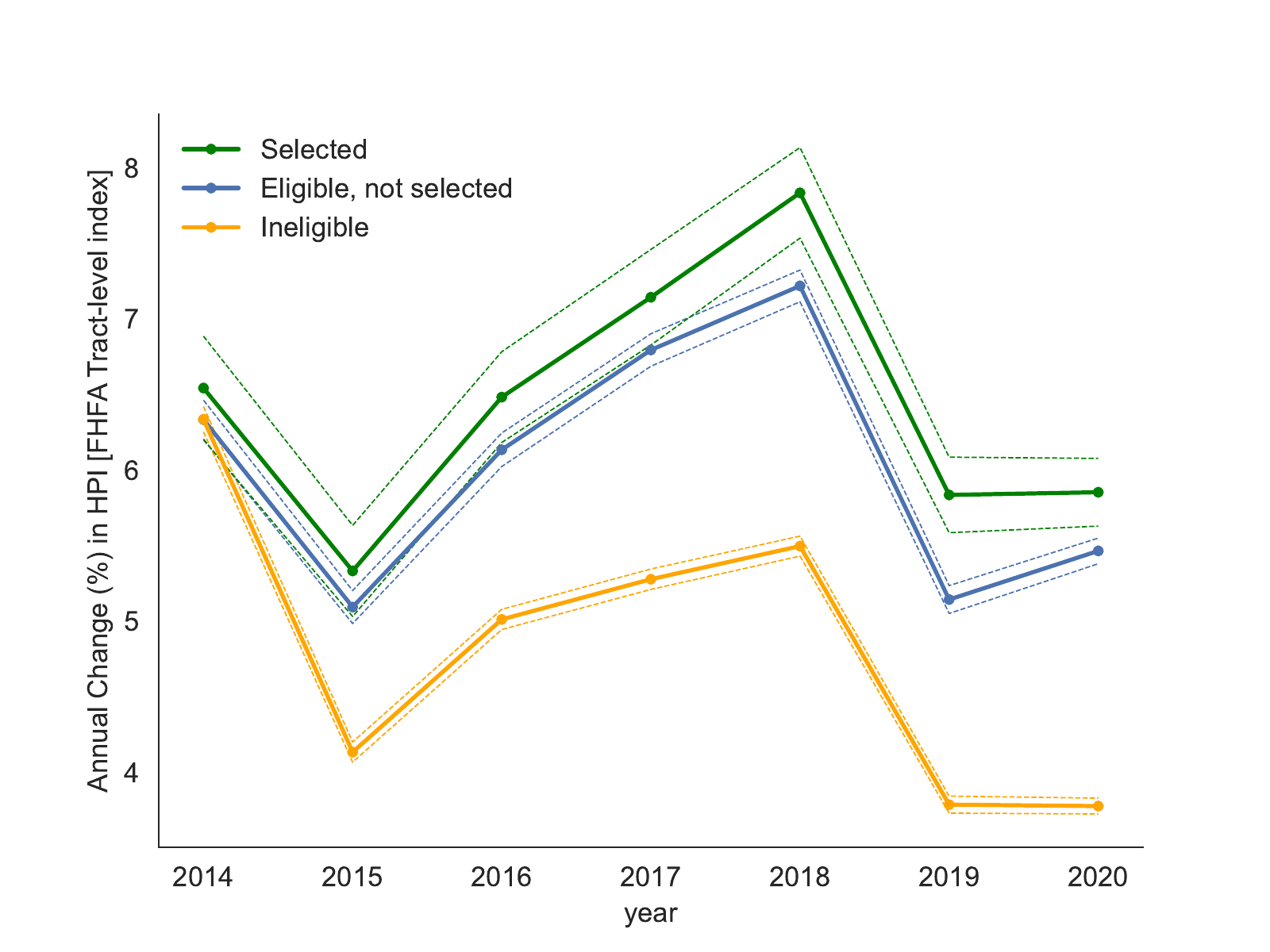}
    
    \includegraphics[width=\textwidth]{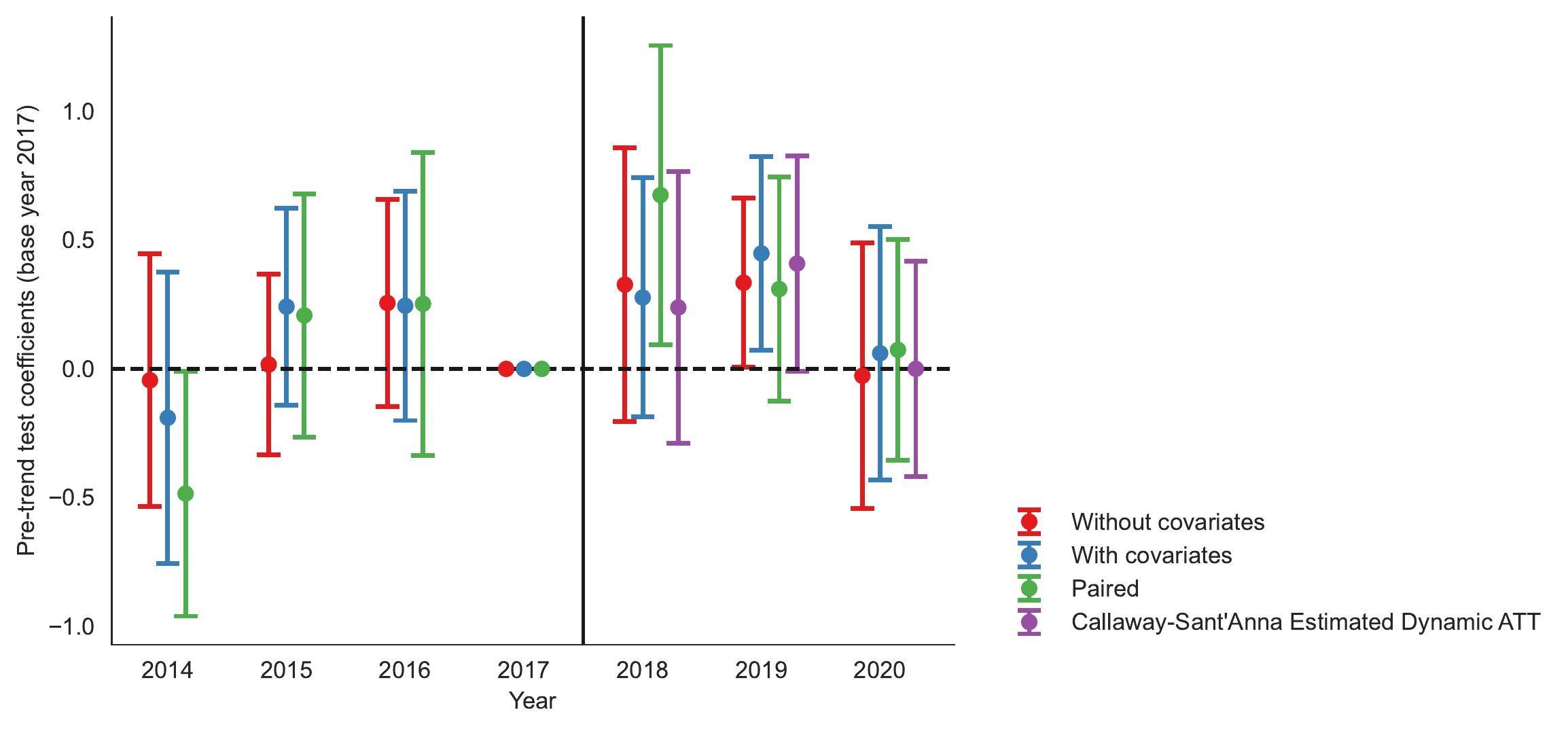}

    \caption{\Cref{fig:zillow_event_study,fig:zillow_event_study} with all eligible
    tracts as control group}
    \label{fig:appendix_fig}
\end{figure}

In the main text, we compare treated zones to control zones that are
strictly low-income. We remove this restriction in this section. We
regenerate \cref{fig:zillow_event_study,fig:zillow_event_study} with 
\cref{fig:appendix_fig}, \cref{tab:tract_and_zip} with 
\cref{tab:tract_and_zip_lic_only_false}, and 
\cref{tab:hetero_effects_} with 
\cref{tab:hetero_effects__lic_only_false}. The
modification does not change our qualitative results. 

\afterpage{\begin{landscape}
    \begin{table}[tbh]
        \caption{Variable definitions, ACS codes, descriptions, and transformations}
        \label{tab:defs}
        \centering
        \vspace{1em}
        \begin{tabular}{lc}
\toprule
{} &                                                            \texttt{description} \\
{} &                                                 \hypertarget{tabcol:defs1}{(1)} \\
\midrule
\texttt{B01003\_001E}                   &                                                             \texttt{population} \\
\texttt{B02001\_002E}                   &                                                      \texttt{white\_population} \\
\texttt{C24020\_001E}                   &                                                   \texttt{employed\_population} \\
\texttt{B08131\_001E}                   &                                                       \texttt{minutes\_commute} \\
\texttt{B09010\_002E}                   &                                                   \texttt{supplemental\_income} \\
\texttt{B15003\_021E}                   &                                                              \texttt{associate} \\
\texttt{B15003\_022E}                   &                                                               \texttt{bachelor} \\
\texttt{B15003\_023E}                   &                                                                 \texttt{master} \\
\texttt{B15003\_024E}                   &                                                   \texttt{professional\_school} \\
\texttt{B15003\_025E}                   &                                                               \texttt{doctoral} \\
\texttt{B16009\_002E}                   &                                                                \texttt{poverty} \\
\texttt{B18140\_001E}                   &                                                       \texttt{median\_earnings} \\
\texttt{B19019\_001E}                   &                                              \texttt{median\_household\_income} \\
\texttt{B25011\_001E}                   &                                                         \texttt{total\_housing} \\
\texttt{B25011\_026E}                   &                                                       \texttt{renter\_occupied} \\
\texttt{B25031\_001E}                   &                                                    \texttt{median\_gross\_rent} \\
\texttt{B27020\_002E}                   &                                                           \texttt{native\_born} \\
\texttt{B27020\_003E}                   &                                              \texttt{native\_born\_hc\_covered} \\
\texttt{pct\_white}                     &                                         \texttt{white\_population / population} \\
\texttt{minutes\_commute}               &                                \texttt{minutes\_commute / employed\_population} \\
\texttt{pct\_higher\_ed}                &  \texttt{(associate + bachelor + professional\_school + doctoral) / population} \\
\texttt{pct\_rent}                      &                                      \texttt{renter\_occupied / total\_housing} \\
\texttt{pct\_native\_hc\_covered}       &                               \texttt{native\_born\_hc\_covered / native\_born} \\
\texttt{pct\_poverty}                   &                                                   \texttt{poverty / population} \\
\texttt{log\_median\_earnings}          &                                                  \texttt{log(median\_earnings)} \\
\texttt{log\_median\_household\_income} &                                         \texttt{log(median\_household\_income)} \\
\texttt{log\_median\_gross\_rent}       &                                               \texttt{log(median\_gross\_rent)} \\
\texttt{pct\_supplemental\_income}      &                                      \texttt{supplemental\_income / population} \\
\texttt{pct\_employed}                  &                                      \texttt{employed\_population / population} \\
\bottomrule
\end{tabular}

        \end{table}
\end{landscape}}

        \begin{table}[tbh]
        \caption{Balance of selected opportunity zones and eligible census tracts}
        \label{tab:balance_of_selected_opportunity_zones_and_eligible_census_tracts}
        \scriptsize
        \centering
        \vspace{1em}
        \begin{threeparttable}
        \begin{tabular}{lccccccc}
\toprule
{} & \multicolumn{2}{c}{Mean} &                                                                                       Diff. & \multicolumn{2}{c}{SE} &                                                                                         $t$ \\
{} &                                                                                Not Selected & \multicolumn{2}{l}{Selected} &                                                                                Not Selected & \multicolumn{2}{l}{Selected} \\
{} & \hypertarget{tabcol:balance_of_selected_opportunity_zones_and_eligible_census_tracts1}{(1)} & \hypertarget{tabcol:balance_of_selected_opportunity_zones_and_eligible_census_tracts2}{(2)} & \hypertarget{tabcol:balance_of_selected_opportunity_zones_and_eligible_census_tracts3}{(3)} & \hypertarget{tabcol:balance_of_selected_opportunity_zones_and_eligible_census_tracts4}{(4)} & \hypertarget{tabcol:balance_of_selected_opportunity_zones_and_eligible_census_tracts5}{(5)} & \hypertarget{tabcol:balance_of_selected_opportunity_zones_and_eligible_census_tracts6}{(6)} \\
\midrule
Population              &                                                                                  $4084.088$ &                                                                                  $4018.494$ &                                                                                   $-65.594$ &                                                                                    $12.424$ &                                                                                    $22.582$ &                                                                                    $-2.545$ \\
Employed pop.           &                                                                                  $1197.027$ &                                                                                  $1085.639$ &                                                                                  $-111.388$ &                                                                                     $4.288$ &                                                                                     $7.356$ &                                                                                   $-13.082$ \\
Avg. commute (min)      &                                                                                    $37.864$ &                                                                                    $37.190$ &                                                                                    $-0.675$ &                                                                                     $0.135$ &                                                                                     $0.223$ &                                                                                    $-2.590$ \\
Median household income &                                                                                 $26076.123$ &                                                                                 $24150.852$ &                                                                                 $-1925.272$ &                                                                                    $47.183$ &                                                                                    $81.187$ &                                                                                   $-20.503$ \\
Median earnings         &                                                                                 $41606.969$ &                                                                                 $36041.271$ &                                                                                 $-5565.698$ &                                                                                    $86.632$ &                                                                                   $145.624$ &                                                                                   $-32.847$ \\
Total housing           &                                                                                  $1478.728$ &                                                                                  $1458.415$ &                                                                                   $-20.313$ &                                                                                     $4.352$ &                                                                                     $7.820$ &                                                                                    $-2.270$ \\
Median gross rent       &                                                                                   $897.552$ &                                                                                   $822.828$ &                                                                                   $-74.724$ &                                                                                     $1.953$ &                                                                                     $3.053$ &                                                                                   $-20.616$ \\
\% White                &                                                                                     $0.624$ &                                                                                     $0.568$ &                                                                                    $-0.057$ &                                                                                     $0.002$ &                                                                                     $0.003$ &                                                                                   $-14.558$ \\
\% Higher ed.           &                                                                                     $0.144$ &                                                                                     $0.129$ &                                                                                    $-0.014$ &                                                                      $4.790 \times 10^{-4}$ &                                                                      $7.741 \times 10^{-4}$ &                                                                                   $-15.677$ \\
\% Rent                 &                                                                                     $0.490$ &                                                                                     $0.557$ &                                                                                     $0.067$ &                                                                                     $0.001$ &                                                                                     $0.003$ &                                                                                    $22.981$ \\
\% Healthcare           &                                                                                     $0.886$ &                                                                                     $0.878$ &                                                                                    $-0.007$ &                                                                      $3.976 \times 10^{-4}$ &                                                                      $7.260 \times 10^{-4}$ &                                                                                    $-9.043$ \\
\% Poverty              &                                                                                     $0.207$ &                                                                                     $0.249$ &                                                                                     $0.043$ &                                                                      $6.371 \times 10^{-4}$ &                                                                                     $0.001$ &                                                                                    $30.347$ \\
\% Supplemental income  &                                                                                     $0.101$ &                                                                                     $0.120$ &                                                                                     $0.019$ &                                                                      $4.143 \times 10^{-4}$ &                                                                      $8.284 \times 10^{-4}$ &                                                                                    $20.923$ \\
\% Employed             &                                                                                     $0.290$ &                                                                                     $0.266$ &                                                                                    $-0.024$ &                                                                      $5.056 \times 10^{-4}$ &                                                                      $8.789 \times 10^{-4}$ &                                                                                   $-23.720$ \\
\bottomrule
\end{tabular}

        \begin{tablenotes}
        \footnotesize
        \item     ``Not Selected'' refers to eligible but not selected opportunity zones.
    Difference is selected minus not selected. Two-sample $t$-statistic reported.
        \end{tablenotes}
        \end{threeparttable}

        \end{table}

\begin{table}[tbh]
        \caption{Covariate balance between geographical pairs (treated minus untreated)}
        \label{tab:paired_balance}
        \centering
        \vspace{1em}
        \begin{tabular}{lcccc}
\toprule
{} &                                       $N$ &                                      Mean &                             Standard Err. &                             $t$-statistic \\
{} & \hypertarget{tabcol:paired_balance1}{(1)} & \hypertarget{tabcol:paired_balance2}{(2)} & \hypertarget{tabcol:paired_balance3}{(3)} & \hypertarget{tabcol:paired_balance4}{(4)} \\
\midrule
Population              &                                   $7,814$ &                                $-563.655$ &                                  $28.437$ &                                 $-19.821$ \\
Employed pop.           &                                   $7,814$ &                                $-389.183$ &                                   $9.982$ &                                 $-38.988$ \\
Avg. commute (min)      &                                   $3,278$ &                                   $2.104$ &                                   $0.229$ &                                   $9.202$ \\
Median household income &                                   $7,796$ &                               $-6989.912$ &                                 $120.019$ &                                 $-58.240$ \\
Median earnings         &                                   $7,800$ &                              $-17255.649$ &                                 $232.043$ &                                 $-74.364$ \\
Total housing           &                                   $7,814$ &                                $-240.748$ &                                  $10.283$ &                                 $-23.413$ \\
Median gross rent       &                                   $7,739$ &                                $-131.982$ &                                   $3.101$ &                                 $-42.562$ \\
\% White                &                                   $7,814$ &                                  $-0.134$ &                                   $0.003$ &                                 $-49.533$ \\
\% Higher ed.           &                                   $7,814$ &                                  $-0.060$ &                                   $0.001$ &                                 $-56.080$ \\
\% Rent                 &                                   $7,808$ &                                   $0.168$ &                                   $0.003$ &                                  $66.720$ \\
\% Healthcare           &                                   $7,813$ &                                  $-0.027$ &                    $6.906 \times 10^{-4}$ &                                 $-39.012$ \\
\% Poverty              &                                   $7,814$ &                                   $0.097$ &                                   $0.001$ &                                  $71.050$ \\
\% Supplemental income  &                                   $7,814$ &                                   $0.048$ &                    $8.911 \times 10^{-4}$ &                                  $53.717$ \\
\% Employed             &                                   $7,814$ &                                  $-0.058$ &                                   $0.001$ &                                 $-55.363$ \\
\bottomrule
\end{tabular}

        \end{table}  %

\afterpage{\begin{landscape}
    
        \begin{table}[tbh]
        \caption{Estimation of ATT using FHFA Tract and ZIP-level data}
        \label{tab:tract_and_zip_lic_only_false}
        \scriptsize
        \centering
        \vspace{1em}
        \begin{threeparttable}
        \begin{tabular}{lcccccc}
\toprule
{} &                                                    TWFE &                                                   TWFE  &                                            Weighting CS &                                            Weighting DR &                                                  Paired &                                   Paired (Linear Trend) \\
{} & \hypertarget{tabcol:tract_and_zip_lic_only_false1}{(1)} & \hypertarget{tabcol:tract_and_zip_lic_only_false2}{(2)} & \hypertarget{tabcol:tract_and_zip_lic_only_false3}{(3)} & \hypertarget{tabcol:tract_and_zip_lic_only_false4}{(4)} & \hypertarget{tabcol:tract_and_zip_lic_only_false5}{(5)} & \hypertarget{tabcol:tract_and_zip_lic_only_false6}{(6)} \\
\midrule
\textbf{Tract-level data}        &                                                         &                                                         &                                                         &                                                         &                                                         &                                                         \\
\quad $\hat\tau$                 &                             $0.155$ [$-0.242$, $0.551$] &                             $0.188$ [$-0.209$, $0.584$] &                               $0.216$ $[-0.160, 0.592]$ &                              $-0.000$ $[-0.386, 0.385]$ &                                $0.359$ $[0.194, 0.523]$ &                               $0.285$ $[-0.226, 0.796]$ \\
\quad                            &                                               ($0.202$) &                                               ($0.154$) &                                               ($0.192$) &                                               ($0.197$) &                                               ($0.084$) &                                               ($0.261$) \\
\quad $p$-value                  &                                                 $0.448$ &                                                 $0.228$ &                                                 $0.260$ &                                                 $0.998$ &                                  $1.900 \times 10^{-5}$ &                                                 $0.274$ \\
\quad Pre-trend test $p$-value   &                                                 $0.533$ &                                                 $0.182$ &                                                 $0.397$ &                                                     --- &                                                 $0.010$ &                                                     --- \\
\quad $(N_1, N_0)$               &                                           (3055, 18468) &                                           (3055, 18468) &                                           (3055, 18468) &                                           (3234, 19148) &                                            (2867, 2867) &                                            (2867, 2867) \\
\quad Covariates                 &                                                      No &                                                     Yes &                                                     Yes &                                                     Yes &                                                     Yes &                                                     Yes \\
\quad Sample                     &                                   Balanced (2014--2020) &                                   Balanced (2014--2020) &                                   Balanced (2014--2020) &                                   Balanced (2017--2020) &                                   Balanced (2014--2020) &                                   Balanced (2014--2020) \\
\midrule \textbf{ZIP-level data} &                                                         &                                                         &                                                         &                                                         &                                                         &                                                         \\
\quad $\hat\tau$                 &                              $1.000$ [$0.641$, $1.360$] &                              $0.364$ [$0.004$, $0.723$] &                                                         &                               $0.038$ $[-0.480, 0.556]$ &                                                         &                                                         \\
\quad                            &                                               ($0.184$) &                                               ($0.195$) &                                                         &                                               ($0.264$) &                                                         &                                                         \\
\quad $p$-value                  &                                  $5.107 \times 10^{-8}$ &                                                 $0.063$ &                                                         &                                                 $0.885$ &                                                         &                                                         \\
\quad Pre-trend test $p$-value   &                                                 $0.039$ &                                                 $0.675$ &                                                         &                                                     --- &                                                         &                                                         \\
\quad $(N_1, N_0)$               &                                            (6105, 8223) &                                            (6105, 8223) &                                                         &                                           (1505, 11356) &                                                         &                                                         \\
\quad Covariates                 &                                                      No &                                                     Yes &                                                         &                                                     Yes &                                                         &                                                         \\
\quad Sample                     &                                   Balanced (2014--2020) &                                   Balanced (2014--2020) &                                                         &                                   Balanced (2017--2020) &                                                         &                                                         \\
\bottomrule
\end{tabular}

        \begin{tablenotes}
        \footnotesize
        \item     \begin{enumerate}

    \item Standard errors are in parenthesis and 95\% confidence intervals are in square brackets.
    Standard errors are clustered at the state level for the tract-level analysis (top panel)
    and clustered at the ZIP level for the ZIP-level analysis (bottom panel).
    Clustering the top panel at the tract level does not qualitatively change results.

    \item Covariates include log median household income, total housing units, percent white,
    percent with post-secondary education,
    percent rental units, percent covered by health insurance among native-born individuals,
    percent below poverty line, percent receiving supplemental income, and percent employed.
    For Column (2), only including log median household income and
percent white as covariates gives $-0.147$ ($0.262$) for the top panel and
$-0.023$ ($0.190$) for the bottom panel.

    \item Pretest for Column (2) interacts covariates with time dummies.

    \item Years 2018 through 2020 are mean-aggregated in Column (4) since the doubly-robust estimation
    only handles two periods.

    \item Discrete treatment in Column (4) is defined as
    the highest 88.0\% of treated
    tract coverage, so as to keep the percentage of treated ZIPs the same as treated tracts.

    \end{enumerate}
    
        \end{tablenotes}
        \end{threeparttable}

        \end{table}
        
\end{landscape}
}

\section{Heterogeneous price effects at the ZIP level}
\label{asec:zip_hetero}

In a previous version of the analysis, we consider the price
effect by residential population analysis aggregated at the ZIP code level
using the Census Business Patterns data for employment, much as we do in
\cref{tab:hetero_effects_}. We reproduce the results below in 
\cref{tab:hetero_effects_zip_} and 
\cref{tab:hetero_effects__lic_only_false}.

        \begin{table}[tbh]
        \caption{Heterogeneous treatment effect by residential population}
        \label{tab:hetero_effects_zip_}
        \scriptsize
        \centering
        \vspace{1em}
        \begin{threeparttable}
        \begin{tabular}{lccc}
\toprule
{} &                                  No Covariates &                                 Few Covariates &                                 All Covariates \\
{} & \hypertarget{tabcol:hetero_effects_zip_1}{(1)} & \hypertarget{tabcol:hetero_effects_zip_2}{(2)} & \hypertarget{tabcol:hetero_effects_zip_3}{(3)} \\
\midrule
Treatment $\times$ Post                      &                     $1.879$ [$1.391$, $2.367$] &                    $0.033$ [$-0.471$, $0.537$] &                    $0.419$ [$-0.072$, $0.909$] \\
                                             &                                      ($0.249$) &                                      ($0.257$) &                                      ($0.250$) \\
Treatment $\times$ Post $\times$ Residential &                  $-1.293$ [$-2.071$, $-0.515$] &                   $-0.686$ [$-1.463$, $0.091$] &                   $-0.342$ [$-1.104$, $0.420$] \\
                                             &                                      ($0.397$) &                                      ($0.396$) &                                      ($0.389$) \\
Pretest $p$-value                            &                         $6.379 \times 10^{-4}$ &                                        $0.439$ &                                        $0.795$ \\
\bottomrule
\end{tabular}

        \begin{tablenotes}
        \footnotesize
        \item     \begin{enumerate}

     \item The table reports the regression \[
    Y_{it}^{\obs} = \mu_i +   \alpha_{it} + \tau_0 \one(t\ge t_0, D_i=1) +
    \tau_1
    \one
    (t\ge t_0, D_i=1, R_i = 1) + \gamma  \one(t\ge t_0, R_i = 1)
    \]
    and Treatment $\times$ Post reports $\tau_0$, while Treatment $\times$ Post
    $\times$ Residential reports $\tau_1$. Here $\alpha_{it} = \alpha_t$ in the
    no-covariate specification and $\alpha_{it} = \alpha_{t}'X_i$ in the covariate
    specification. $R_i$ is an indicator for whether the employment to
    residential population ratio is lower than median.

    \item Standard errors are in parenthesis and 95\% confidence intervals are in square brackets.
    Standard errors are clustered at the ZIP level.

    \item ``All covariates'' consists of log median household income, total housing units, percent white,
    percent with post-secondary education,
    percent rental units, percent covered by health insurance among native-born individuals,
    percent below poverty line, percent receiving supplemental income, and percent employed. ``Few covariates''
    consists of only log median household income and total housing units.
    \end{enumerate}
    
        \end{tablenotes}
        \end{threeparttable}

        \end{table}

        \begin{table}[tbh]
        \caption{Heterogeneous treatment effect by residential population}
        \label{tab:hetero_effects__lic_only_false}
        \scriptsize
        \centering
        \vspace{1em}
        \begin{threeparttable}
        \begin{tabular}{lccc}
\toprule
{} &                                             No Covariates &                                            Few Covariates &                                            All Covariates \\
{} & \hypertarget{tabcol:hetero_effects__lic_only_false1}{(1)} & \hypertarget{tabcol:hetero_effects__lic_only_false2}{(2)} & \hypertarget{tabcol:hetero_effects__lic_only_false3}{(3)} \\
\midrule
Treatment $\times$ Post                      &                                $1.752$ [$1.314$, $2.190$] &                               $0.068$ [$-0.384$, $0.520$] &                                $0.502$ [$0.061$, $0.943$] \\
                                             &                                                 ($0.223$) &                                                 ($0.231$) &                                                 ($0.225$) \\
Treatment $\times$ Post $\times$ Residential &                             $-1.464$ [$-2.214$, $-0.714$] &                             $-0.993$ [$-1.743$, $-0.244$] &                              $-0.551$ [$-1.289$, $0.188$] \\
                                             &                                                 ($0.383$) &                                                 ($0.382$) &                                                 ($0.377$) \\
Pretest $p$-value                            &                                    $9.489 \times 10^{-5}$ &                                                   $0.060$ &                                                   $0.532$ \\
\bottomrule
\end{tabular}

        \begin{tablenotes}
        \footnotesize
        \item     \begin{enumerate}

     \item The table reports the regression \[
    Y_{it}^{\obs} = \mu_i +   \alpha_{it} + \tau_0 \one(t\ge t_0, D_i=1) +
    \tau_1
    \one
    (t\ge t_0, D_i=1, R_i = 1) + \gamma  \one(t\ge t_0, R_i = 1)
    \]
    and Treatment $\times$ Post reports $\tau_0$, while Treatment $\times$ Post
    $\times$ Residential reports $\tau_1$. Here $\alpha_{it} = \alpha_t$ in the
    no-covariate specification and $\alpha_{it} = \alpha_{t}'X_i$ in the covariate
    specification. $R_i$ is an indicator for whether the employment to
    residential population ratio is lower than median.

    \item Standard errors are in parenthesis and 95\% confidence intervals are in square brackets.
    Standard errors are clustered at the ZIP level.

    \item ``All covariates'' consists of log median household income, total housing units, percent white,
    percent with post-secondary education,
    percent rental units, percent covered by health insurance among native-born individuals,
    percent below poverty line, percent receiving supplemental income, and percent employed. ``Few covariates''
    consists of only log median household income and total housing units.
    \end{enumerate}
    
        \end{tablenotes}
        \end{threeparttable}

        \end{table}

\section{Permitting analysis}

\subsection{Decomposing units}
\label{asub:unittype}

\begin{figure}[tb]
    \centering
    \includegraphics[width=\textwidth]
    {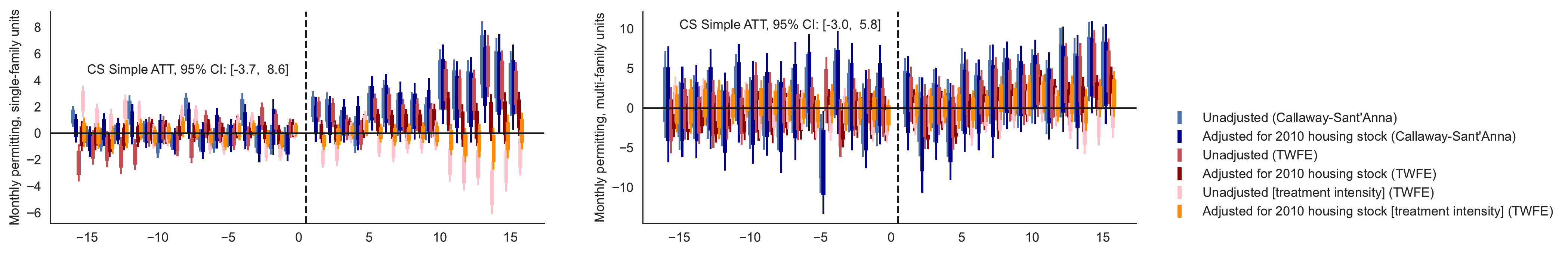}
    \caption{Decomposing permit analysis by unit type}
    \label{fig:unit_type}
\end{figure}

We carry out the analysis in the top-left panel of \cref{fig:quant}, except
we decompose permits for total units into single-family units and
multi-family units. The resulting event study plots are shown in 
\cref{fig:unit_type}. We see that most of our results are driven by
differences in single-family units. 

\subsection{Changes in changes}
\label{asec:cic}

\begin{figure}[tb]
    \centering
    \includegraphics{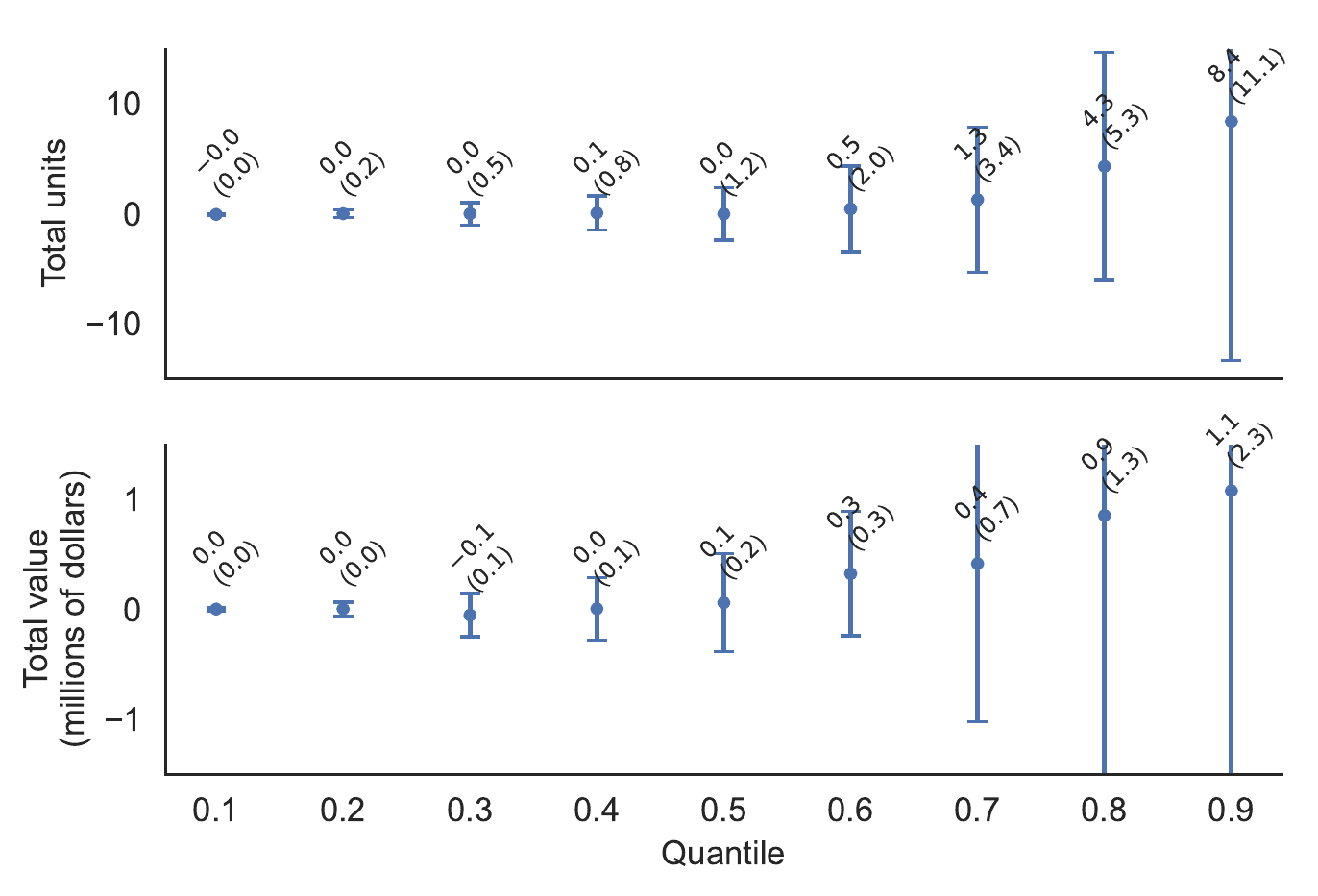}
    \caption{Changes-in-changes estimates of quantile treatment effects
    for permitting results}
    \label{fig:cic}
\end{figure}

In this section we detail how we computed changes-in-changes estimates for
the quantile treatment effects in 
permitting units and value \citep{athey2006identification}. We use the
implementation of \cite{callaway2016quantile}, which is suitable for a
two-period setting. As a result, we engineer a two-period dataset by
aggregating pre-treatment and post-treatment months for each Census place.
The resulting estimates are shown in \cref{fig:cic}.

\end{document}